\documentclass[showpacs,nofootinbib]{revtex4}
\usepackage{graphicx}

\newcommand{\CA}{{\cal A}}

\newcommand{\CL}{{\cal L}}

\newcommand{\CN}{{\cal N}}
\newcommand{\CO}{{\cal O}}

\newcommand{\CR}{{\cal R}}

\newcommand{\bear}{\begin{array}}  \newcommand{\eear}{\end{array}}
\newcommand{\bea}{\begin{eqnarray}}  \newcommand{\eea}{\end{eqnarray}}
\newcommand{\beq}{\begin{equation}}  \newcommand{\eeq}{\end{equation}}
\newcommand{\bef}{\begin{figure}}  \newcommand{\eef}{\end{figure}}
\newcommand{\bec}{\begin{center}}  \newcommand{\eec}{\end{center}}
\newcommand{\non}{\nonumber}  
\newcommand{\lmk}{\left(}  \newcommand{\rmk}{\right)}
\newcommand{\lkk}{\left[}  \newcommand{\rkk}{\right]}
\newcommand{\lhk}{\left \{ }  \newcommand{\rhk}{\right \} }
\newcommand{\del}{\partial}  
\newcommand{\vect}[1]{\mbox{\boldmath${#1}$}}
 
\newcommand{\la}{\left\langle} \newcommand{\ra}{\right\rangle}

\newcommand{\wlambda}{\widetilde{\lambda}}

\begin{document}

\title{Supergravity based inflation models: a review}
\author{Masahide Yamaguchi}
\affiliation{Department of Physics, Tokyo Institute of Technology, Tokyo
152-8551, Japan}
\date{\today}
\begin{abstract}
In this review, we discuss inflation models based on supergravity. After
explaining the difficulties in realizing inflation in the context of
supergravity, we show how to evade such difficulties. Depending on types
of inflation, we give concrete examples, particularly paying attention
to chaotic inflation because the ongoing experiments like Planck might
detect the tensor perturbations in near future. We also discuss
inflation models in Jordan frame supergravity, motivated by Higgs
inflation.
\end{abstract}


\maketitle


\section{Introduction}

\label{sec:introduction} 

Recent observations of the cosmic microwave background (CMB)
anisotropies \cite{WMAP} strongly support the presence of accelerated
expansion era in the early universe, that is, inflation. Although
inflation was originally introduced to solve the horizon, flatness,
monopole problems, and so on \cite{inflation}, it can also give an
explanation to the origin of primordial density fluctuations, which are
responsible for the large scale structure of the present Universe.
Unfortunately, however, the origin of inflation, namely what (which
field) caused inflation, is still unknown until now and it is one of the
most important mysteries of particle physics and cosmology.

Inflation can be caused by the potential energy of a scalar field. Such
a potential must be relatively flat in order to guarantee long duration
of inflation and small deviation of scale invariance of primordial
density fluctuations. However, the flatness of the scalar potential can
be easily destroyed by radiative corrections. One of the leading
theories to protect a scalar field from radiative corrections is
supersymmetry (SUSY), which also gives an attractive solution to the
(similar) hierarchy problem of the standard model (SM) of particle
physics as well as the unification of the three gauge couplings. In
particular, its local version, supergravity, would govern the dynamics
of the early Universe, when high energy physics was important. Thus, it
is quite natural to consider inflation in the framework of
supergravity. However, in fact, it is a non-trivial task to incorporate
inflation in supergravity. This is mainly because a SUSY breaking
potential term, which is indispensable to inflation, generally gives a
would-be inflaton an additional mass, which spoils the flatness of an
inflaton potential
\cite{Hmass,Copeland:1994vg,Stewart:1994ts}. Specifically, the
exponential factor appearing in the F-term is troublesome. Assuming the
canonical K\"ahler potential, this exponential factor generates the
additional mass comparable to the Hubble parameter for a field value
smaller than the reduced Planck scale $M_p \simeq 2.4 \times
10^{18}$~GeV, which makes it difficult to realize small field inflation
such as new and hybrid inflations in supergravity. In addition, it
prevents a scalar field from acquiring a value larger than $M_p$. This
fact implies that it is almost impossible to realize large field
inflation like chaotic inflation in supergravity. In fact, after the
original chaotic inflation had been proposed, almost twenty years passed
until a natural model of chaotic inflation in supergravity appeared. The
main purpose of this review is to explain how to circumvent these
difficulties and how to realize inflation in supergravity.

While the standard model of particle physics includes only one scalar
field (Higgs field), scalar fields are ubiquitous in the supersymmetric
theories. The Higgs field, unfortunately, cannot be responsible for
inflation as long as it has the canonical kinetic term and is minimally
coupled to gravity because it predicts too large density fluctuations as
well as too large tensor-to scalar ratio. Recently, a possibility to
realize Higgs inflation with a non-minimal coupling to gravity
\cite{Bezrukov:2007ep} and/or a nontrivial kinetic term
\cite{Germani:2010gm,Nakayama:2010sk,Kamada:2010qe} has drawn much
attention.\footnote{See Refs. \cite{Murayama:1992ua} for other attempts
to realize inflation in the context of SM and its SUSY extentions.} We
are going to mention such a possibility in the last part of this
review. However, first of all, we concentrate on inflation models with
an (almost) canonical kinetic term and a minimal coupling to gravity.

Finally, we would like to make a comment on the relation between
supergravity and superstring. While supergravity is a {\it field}
theory, actually, the low energy effective field theory of superstring,
superstring is a {\it string} theory though other higher dimensional
objects like D-branes are found to join it. Since inflation only needs
the positive potential energy, it can be realized in the framework of an
(effective) scalar field. Thus, inflation does not necessarily require
superstring theory. However, superstring theory can give us concrete
forms of the potentials like K\"ahler potential, superpotential, and
gauge kinetic function in supergravity.  Therefore, once we find
particular forms of such potentials suitable for inflation, it would be
interesting to investigate whether such forms naturally appear in the
framework of superstring. Since we do not pursue such possibility in
this review, we refer recent reviews of string inflation
\cite{stringreview} as well as other excellent reviews of supergravity
inflation \cite{SUGRAinflation}.

The organization of this review is as follows: In the next section, we
first give the basics of inflation and a scalar potential in
supergravity, which is composed of F-term and D-term. Then we explain
why it is difficult to incorporate inflation in the context of
supergravity. In Sec. III, after giving a general discussion how to
evade such a difficulty, we will give concrete examples of inflation
based on F-term or D-term for each inflation type. In Sec. IV, we will
discuss inflation models in the Jordan frame supergravity. Final section
is devoted to conclusion.

\section{Basic formulae and difficulty in realizing inflation in supergravity}

\label{sec:basic}

\subsection{Basic formulae of inflation dynamics and primordial perturbations}

\label{subsec:formulae}

Let us consider a single-field $\phi$ inflation model with the canonical
kinetic term and potential $V(\phi)$, whose action is given by
\beq
  S = \int d^4x \sqrt{-g} \CL 
    = \int d^4x \sqrt{-g} 
      \biggl[ - \frac12 \del_{\mu}\phi\del^{\mu}\phi  - V(\phi) \biggr].
  \label{eq:lagrangian}
\eeq
Here $g$ is the determinant of the metric $g_{\mu\nu}$ and the scalar
field is assumed to be minimally coupled to gravity. The extention to
multi-field case is straightforward. Provided that the scalar field is
homogeneous and the metric is $ds^2 = - dt^2+a(t)^2 d\vect{x}^2$, the
equation of motion and the Friedmann equations are given by
\bea
 &&  \ddot{\phi} + 3 H \dot{\phi} + V'(\phi) = 0, \non \\
 &&  H^2 = \frac13 \lkk \frac12 \dot{\phi}^2 + V(\phi) \rkk,
\eea
where the dot and the dash represent the derivatives with respect to the
cosmic time and the scalar field $\phi$, respectively. Here and
hereafter we set the reduced Planck scale $M_p$ to be unity unless
otherwise stated. These equations can be approximated as
\bea
 &&  3 H \dot{\phi} + V'(\phi) \simeq 0, \non \\
 &&  H^2 \simeq \frac{V(\phi)}{3},
 \label{eq:sleq}
\eea
as long as the following two slow-roll conditions are satisfied,
\bea
 && \epsilon \equiv \frac12 \lmk \frac{V'}{V} \rmk^2;~~~~
      \epsilon  \ll 1, \label{eq:slowroll1} \\
 && \eta \equiv \frac{V''}{V};~~~~|\eta| \ll 1.
                       \label{eq:slowroll2} 
\eea
From Eqs. (\ref{eq:sleq}), the $e$-folding number $N$ is estimated as
\beq
  N = \int_{t}^{t_e} H dt 
    \simeq \int_{\phi}^{\phi_e} H \frac{d\phi}{\dot{\phi}}  
    \simeq \int_{\phi_e}^{\phi} \frac{V}{V'} d\phi,    
\eeq
where $t_e$ and $\phi_e$ are the cosmic time and the field value at the
end of inflation. Typically, a cosmologically interesting scale
corresponds to $N \sim 50$ or $60$ depending on inflationary energy
scale and cosmic history after inflation.

During inflation, the curvature perturbations were generated through
inflaton fluctuations and then their amplitude in the comoving gauge
$\CR$ \cite{Bardeen:1980kt} on the comoving scale $2\pi/k$ is given by
\beq
 \CR^2(k) = \frac{1}{4\pi^2} \frac{H^4(t_k)}{\dot{\phi}^2(t_k)} 
          \simeq \frac{1}{24 \pi^2} \frac{V}{\epsilon}.
 \label{eq:fluc}
\eeq
Here $t_k$ is the epoch when $k$ mode left the Hubble radius during
inflation \cite{pert}. The spectral index of the curvature perturbation
is calculated as
\beq
  n_s-1 \equiv \frac{d\ln \CR^2(k)}{d\ln k}
        \simeq - 6 \epsilon + 2 \eta.
  \label{eq:index}
\eeq
On the other hand, the tensor perturbations (gravitational wave) $h$ are
also generated and their amplitude on the comoving scale $2\pi/k$ is given
by \cite{Starobinsky:1979ty}
\beq
 h^2(k) = 8 \lmk \frac{H(t_k)}{2\pi} \rmk^2 \simeq \frac{2V}{3\pi^2},
 \label{eq:tensorfluc}
\eeq
where the coefficient $8$ comes from the canonical normalization and the
number of polarization of tensor modes. Then, the tensor to scalar ratio
$r$ is given by
\beq
  r \equiv \frac{h^2(k)}{\CR^2(k)} \simeq 16 \epsilon.
\eeq
Detailed derivation of these standard formulae is given in the textbook
\cite{Lyth:2009zz}, for example.

Recent observations of the Wilkinson Microwave Anisotropy Probe (WMAP)
can strongly constrain these observable quantities as \cite{WMAP}
\bea
 &&  \CR^2(k_0) = 2.441^{+0.088}_{-0.092} \times 10^{-9}, \\
 &&  n_s = 0.963 \pm 0.012, \\
 &&  r < 0.24,
\eea
with $k_0 = 0.002 {\rm Mpc}^{-1}$. Note that, if we allow the running of
the spectral index, these constraints can be significantly relaxed.

\subsection{Difficulty in realizing inflation in supergravity}

\label{subsec:difficulty}

The scalar part of the Lagrangian in supergravity is determined by the
three functions, K\"ahler potential $K(\Phi_i,\Phi^{\ast}_i)$,
superpotential $W(\Phi_i)$, and gauge kinetic function $f(\Phi_i)$
\cite{SUSYreview}. While the last two functions ($W$ and $f$) are
holomorphic functions of complex scalar fields, the first one $K$ is not
holomorphic and a real function of the scalar fields $\Phi_i$ and their
conjugates $\Phi_i^{\ast}$. Notice that the above three functions are
originally the functions of (chiral) superfields. Since we are mainly
interested in the scalar part of a superfield, we identify a superfield
with its complex scalar component and write both of them by the same
letter.

The action of complex scalar fields minimally coupled to gravity
consists of kinetic and potential parts,
\beq
  S = \int d^4x \sqrt{-g} \biggl[
       \frac{1}{\sqrt{-g}} \CL_{\rm kin} - V(\Phi_i,\Phi_i^{\ast}) \biggr].
  \label{eq:lagrangian2}
\eeq
The kinetic terms of the scalar fields are determined by K\"ahler
potential $K$ and given by
\beq
  \frac{1}{\sqrt{-g}} \CL_{\rm kin} = - K_{ij^{\ast}}
                  D_{\mu} \Phi_i D_{\nu} \Phi_j^{\ast} g^{\mu\nu},
  \label{eq:kinetic}
\eeq
where
\beq
  K_{ij^{\ast}} = \frac{\del^2 K}{\del \Phi_i \del \Phi_j^{\ast}},
  \label{eq:Kij}
\eeq
and $D_{\mu}$ represents the gauge covariant derivative. Here and
hereafter, the lower indices of the K\"ahler potential stand for the
derivatives. The potential $V$ of scalar fields $\Phi_i$ is made of two
terms, F-term $V_F$ and D-term $V_D$. The F-term $V_F$ is determined by
superpotential $W$ as well as K\"ahler potential $K$,
\beq
  V_F = e^{K} \left[ D_{\Phi_i}W K_{ij^{\ast}}^{-1} D_{\Phi_j^{*}}W^{*} 
                     - 3 |W|^{2} \right],
  \label{eq:Fpotential}
\eeq
with 
\beq
  D_{\Phi_i}W = \frac{\del W}{\del \Phi_i} 
    + \frac{\del K}{\del \Phi_i}W.
  \label{eq:DW}
\eeq
On the other hand, D-term $V_D$ is related to gauge symmetry and given
by gauge kinetic function as well as K\"ahler potential,
\beq
  V_{D} = \frac12 \sum_{a} \lkk {\rm Re} f_{a}(\Phi_i) \rkk^{-1} 
          g_a^2 D_a^2,
  \label{eq:Dpotential}
\eeq
with
\beq
 D_a = \Phi_i (T_a)^{i}_{j} 
   \frac{\del K}{\del \Phi_{j}} + \xi_a.
\eeq
Here, the subscript $a$ represents a gauge symmetry, $g_a$ is a
gauge coupling constant, and $T_a$ is an associated generator. $\xi_a$
is a so-called Fayet-Iliopoulos (FI) term and can be non-zero only when
the gauge symmetry is Abelian, that is, $U(1)$ symmetry. Notice that
only a combination
\beq
  G \equiv K + \ln |W|^2
\eeq
is physically relevant. Then, the kinetic and the potential terms are
invariant under the following K\"ahler transformation,
\bea
 && K(\Phi_i,\Phi_i^{\ast})  \longrightarrow 
    K(\Phi_i,\Phi_i^{\ast}) - U(\Phi_i) - U^{\ast}(\Phi_i^{\ast}), \non \\
 && W(\Phi_i)  \longrightarrow e^{U(\Phi_i)} W(\Phi_i),
\eea 
where $U(\Phi_i)$ is any holomorphic function of the fields $\Phi_i$.

From the potential form (\ref{eq:Fpotential}), it is manifest that, in
order to acquire the positive energy density necessary for inflation, at
least one of the terms $D_{\Phi_i}W$ must be non-zero. Since these terms
are the order parameters of SUSY, it turned out that inflation is always
accompanied by the SUSY breaking, whose effect could be transmitted to
any scalar field and generate a dangerous mass term. Specifically,
taking (almost) canonical K\"ahler potential,
\beq
  K(\Phi_i,\Phi_i^{\ast}) = \sum_{i} |\Phi_i|^2 + \cdots,
\eeq
the kinetic term of the scalar fields $\Phi_i$ becomes (almost)
canonical. Here the ellipsis stands for higher order terms.  Then, the
F-term potential, $V_F$, is approximated as
\cite{Copeland:1994vg,Stewart:1994ts}
\bea
  V_F &=& \exp{\lmk \sum_{i} |\Phi_i|^2 + \cdots \rmk} \times
          \lhk 
      \lkk \frac{\del W}{\del \Phi_i} + (\Phi_i^{\ast}+\cdots)W \rkk  
      \sum_{i,j} \lmk \delta_{ij} + \cdots \rmk
      \lkk \frac{\del W^{\ast}}{\del \Phi_j^{\ast}} + (\Phi_j+\cdots)W^{\ast} \rkk  
      - 3 |W|^2 \rhk \non \\
      &=& V_{\rm global} + V_{\rm global} \sum_{i} |\Phi_i|^2 
          + {\rm other~terms},
\eea
where $V_{\rm global}$ is the effective potential in the global SUSY
limit and given by
\beq
  V_{\rm global} = 
    \sum_i \left| \frac{\del W}{\del \Phi_i} \right|^2.
\eeq
Thus, {\it any} scalar field including a would-be inflaton receives the
effective mass squared $V_{\rm global} = 3 H^2$, which gives a
contribution of order unity to the slow-roll parameter $\eta$ and breaks
one of the slow-roll conditions necessary for successful inflation,
\beq
  \eta = \frac{V''}{V} = 1 + {\rm other~terms}. 
\eeq
This is the main difficulty in incorporating inflation in supergravity
and is called the $\eta$ problem.\footnote{If inflation can be realized
in the strongly dissipative system \cite{Berera:1995ie}, which we do not
deal with in this review, the large effective masses coming from the
supergravity corrections would be harmless in such a dynamical system
\cite{BasteroGil:2006vr}.}

Several methods have been proposed to evade this problem thus far :

\begin{itemize}

\item

Use K\"ahler potential different from (almost) canonical one. In this
case, we have two possibilities to obtain flat potentials, both of which
are related. When the K\"ahler potential is far from canonical, so is
the kinetic term of the scalar field. By redefining the scalar field
such that its kinetic term is canonically normalized, the effective
potential could be flat even if it was originally steep. In the next
section, we mainly focus on (almost) canonical K\"ahler potential and
consider this possibility only in chaotic inflation of F-term models.

The second possibility is to imposes some conditions (or symmetries) on
the K\"ahler potential and/or the superpotential, which guarantee the
flatness of the potential. For example, Heisenberg symmetry was used to
avoid the additional scalar mass \cite{Gaillard:1995az}. Another
condition is given in Ref. \cite{Stewart:1994ts} and, interestingly, the
required form of the K\"ahler potential appears in weakly coupled string
theory. Therefore, it is better to discuss such possibilities in the
context of string theory and hence we skip them in this review.

\item

Use quantum corrections. The potential given in
Eq. (\ref{eq:Fpotential}) is a classical one. In the case that an
inflaton has (large) Yukawa and/or gauge interactions, quantum
corrections significantly modify the potential so that the inflaton mass
runs with scale \cite{Stewart:1996ey}. We mention this possibility in
hybrid inflation of F-term models.

\item

Use a special form of superpotential with (almost) canonical K\"ahler
potential. Actually, as shown below, in the case that superpotential is
linear in an inflaton, the inflaton effective mass becomes negligible so
that the $\eta$ problem is avoided
\cite{Copeland:1994vg,Stewart:1994ts,Kumekawa:1994gx}. Such a form of
superpotential can be easily realized by imposing the $R$ symmetry with
the $R$ charge of the inflaton to be two and the others to be zeroes.
In the next section, we will give concrete examples of inflation with
such types of superpotential.

\item

Use D-term potential. The $\eta$ problem is peculiar to F-term
potential. Therefore, if we can get the positive energy in D-term
potential, inflation is easily realized \cite{Stewart:1994ts}. In the
next section, we will give concrete examples of such D-term inflation
models.

\end{itemize}

Before closing this section, we note that the solutions to the $\eta$
problem is not sufficient for large field inflation such as chaotic and
topological inflations.  Though they need the field value of an inflaton
larger than unity, the exponential factor of F-term potential prevents
the inflaton from taking such a large value as long as the K\"ahler
potential is (almost) canonical. Thus, we need another prescription to
realize large field inflation, which will be given later in the
corresponding models.

\section{Inflation models in supergravity}

In this section, we give concrete examples of successful inflation model
for each type. In the former subsection, inflation models supported by
F-term potential are given, and in the latter subsection, D-term models
will be discussed. Recent observations are so precise that the original
models may be disfavored in some cases and hence some extended models
are proposed. In addition, the gravitino problem is another important
issue in constructing inflation models in supergravity. The gravitino is
a superpartner of graviton and is a fermion with spin $3/2$. During the
reheating stage of inflation, gravitinos are copiously produced so that
they may easily destroy light elements synthesized during the big bang
nucleosynthesis (BBN) or overclose the Universe, depending on its mass
and lifetime determined by the SUSY breaking mechanism. Recently, new
mechanism to produce gravitinos during reheating
\cite{Kawasaki:2006gs,Endo:2006qk} was found in addition to the
conventional production mechanism from thermal plasma
\cite{Kawasaki:2004yh}. Thus, strong constraint on reheating temperature
is imposed on inflation models, which may also rule out the original
models for some range of gravitino masses. However, in this review, we
stick to the original or the simplest models simply because one can
easily understand the essence of each model. See the references for more
elaborated models to fit the observed results well and to avoid the
gravitino problem.

\label{sec:models}

\subsection{F-term inflation}

\label{subsec:F-term}

As discussed in the previous section, the exponential factor appearing
in F-term potential can give a would-be inflaton an additional mass and
rule it out as an inflaton. One of the methods to circumvent this
difficulty is to adopt superpotential linear in the inflaton $\Phi$
\cite{Copeland:1994vg,Stewart:1994ts,Kumekawa:1994gx},
\beq
  W = \Phi f(\chi_i),
\eeq
where $\chi_i$ is a field other than the inflaton and $f$ is a
holomorphic function of $\chi_i$. This type of superpotential can be
easily realized by imposing the $R$ symmetry, under which they are
transformed as $\Phi(\theta) \longrightarrow e^{2i\alpha}\Phi(\theta
e^{i\alpha})$, $f(\chi_i)(\theta) \longrightarrow f(\chi_i)(\theta
e^{i\alpha})$. Notice that the canonical K\"ahler potential given by
\beq
  K = |\Phi|^2 + \sum_i \left| \chi_i \right|^2
\eeq
respects this $R$ symmetry. In this case, the F-term potential is given
and approximated as
\bea
  V_F &=& e^K \lkk |f|^2 \lmk 1 - |\Phi|^2 + |\Phi|^4 \rmk 
          + |\Phi|^2 \left| \frac{\del f}{\del \chi_i} + \chi_i^{\ast} f
                     \right|^2 \rkk
          \non \\
      &\simeq& V_0 \lmk 1 + \frac{|\Phi|^4}{2} + |\chi_i|^2 \rmk 
          + |\Phi|^2 \left| \frac{\del f}{\del \chi_i} + \chi_i^{\ast} f
                     \right|^2 , 
\eea
where $V_0 \equiv |f|^2 = |\del W/\del \Phi|^2$ and we have expanded the
exponential factor for $|\Phi|, |\chi_i| \ll 1$. It is found that there
is no inflaton mass associated with $V_0$, while the other fields
$\chi_i$ acquire the additional masses squared $V_0 \simeq 3 H^2$, which
usually drive $\chi_i$ to zeros. Though the second term in the right
hand side of the last equation is the mass term of the inflaton, it is
typically very small and hence the $\eta$ problem is evaded. In
particular, in case that every $\chi_i$ goes to zero and $\del f/\del
\chi_i = 0$ at the origin of $\chi_i$, the inflaton is exactly massless.

Now, we are ready to give concrete examples of successful inflation
models for each type, that is, new inflation, hybrid inflation, chaotic
inflation, and topological inflation, though additional tricks are
necessary for the last two types.

\subsubsection{New inflation}

\label{subsubsec:Fnew}

New inflation was proposed as the first slow-rolling inflation model
\cite{new}.  As a concrete model of new inflation, which uses the F-term
potential, we consider the model proposed by Izawa and Yanagida
\cite{Izawa:1996dv}.

A chiral superfield $\Phi$ with the $R$ charge $2/(n+1)$ is introduced
as an inflaton. In this model, the U$(1)_R$ symmetry is assumed to be
dynamically broken to a discrete $Z_{2n}$ $R$ symmetry at a scale $v \ll
1$. Then, the superpotential is given by
\beq
 W = v^2\Phi - \frac{g}{n+1}\Phi^{n+1}, 
\label{eq:newsuper}
\eeq
where $g$ is a coupling constant of order unity. We assume that both $g$
and $v$ are real and positive in addition to $n \ge 3$, for
simplicity. The $R$-invariant K\"ahler potential is given by
\beq
 K=|\Phi|^2+\cdots, 
\label{eq:newkahler}
\eeq
where the ellipsis stands for higher order terms, which we ignore for a
while.

The scalar potential is obtained from Eqs. (\ref{eq:newsuper}) and
(\ref{eq:newkahler}) by use of the formula given in
Eq. (\ref{eq:Fpotential}) and reads
\bea
  V(\Phi) &=& e^{|\Phi|^2} 
              \lkk~ \left| \lmk 1+|\Phi|^2 \rmk v^2
             -\lmk 1+\frac{|\Phi|^2}{n+1} \rmk g\Phi^n \right|^2 
             - 3 |\Phi|^2 \left| v^2 - \frac{g}{n+1}\Phi^n \right|^2\rkk .
\eea
It has a minimum at
\beq
 |\Phi|_{\rm min} \simeq \lmk\frac{v^2}{g}\rmk^{\frac{1}{n}}
 ~~~~{\rm and~~~Im~}\Phi^n_{\rm min}=0, 
\eeq
with negative energy density
\beq
 V \lmk \Phi_{\rm min} \rmk \simeq 
            - 3e^{|\Phi|^2} \left| W \lmk \Phi_{\rm min} \rmk \right|^2
                            \simeq 
            -3 \lmk \frac{n}{n+1} \rmk^2 v^4 \left| \Phi_{\rm min} \right|^2.
\eeq
You may wonder if this negative value may be troublesome. However, in
the context of SUSY, we may interpret that such negative potential
energy is almost canceled by positive contribution due to the
supersymmetry breaking, $\Lambda_{\rm SUSY}^4$, and that the residual
positive energy density is responsible for the present dark
energy. Then, we can relate the energy scale of this model with the
gravitino mass $m_{3/2}$ as
\beq
 m_{3/2} \simeq \frac{n}{n+1} \lmk \frac{v^2}{g} \rmk^{\frac{1}{n}} v^2.
\eeq

Identifying the real part of $\Phi$ with the inflaton $\phi \equiv
\sqrt{2} {\rm Re} \Phi$, the dynamics of the inflaton is governed by the
following potential,
\beq
 V(\phi) \simeq v^4 - \frac{2g}{2^{n/2}}v^2 \phi^n 
  + \frac{g^2}{2^n}\phi^{2n}. 
\label{eq:neweffpot}
\eeq
You can easily find that the first constant term dominates the potential
energy and that the last term is negligible during inflation. Then, the
Hubble parameter during inflation is given by $H=v^2/\sqrt{3}$. On the
other hand, the slow-roll equation of motion reads
\beq
 3 H \dot{\phi} \simeq -V'(\phi) 
   = \frac{ng}{2^{\frac{n-2}{2}}} v^2 \phi^{n-1},
\label{eq:neweqm}
\eeq
and the slow-roll parameters are given by
\beq
 \epsilon \simeq \frac{n^2g^2}{2^{n-1}} \frac{\phi^{2(n-1)}}{v^4},~~~
 \eta \simeq - \frac{n(n-1)g}{2^{\frac{n-2}{2}}} \frac{\phi^{n-2}}{v^2}.
\eeq
Thus, inflation lasts as long as $\phi$ is small enough, and it ends
at
\beq
  \phi=\sqrt{2} \lmk \frac{v^2}{gn(n-1)} \rmk^{\frac{1}{n-2}} \equiv
  \phi_e,
\eeq
when $|\eta|$ becomes as large as unity. The $e$-folding number of new
inflation is estimated as
\beq
 N = \int_{\phi_N}^{\phi_e} 
        \frac{2^{\frac{n-2}{2}}v^2}{ng\phi^{n-1}}d\phi
   = \frac{2^{\frac{n-2}{2}}v^2}{ng(n-2)} \frac{1}{\phi_N^{n-2}}
     -\frac{n-1}{n-2},
\label{eq:nnzero}
\eeq
where $\phi_N$ is a field value corresponding to the $e$-folds equal to
$N$. The amplitude and the spectral index of primordial density
fluctuations are expressed in terms of $N$ as
\bea
  \CR^2 &=& \frac{1}{24 \pi^2} \frac{V}{\epsilon}
        \simeq \frac{2^{n-4}}{3\pi^2 n^2 g^2} \frac{v^8}{\phi_N^{2(n-1)}}
        \simeq \frac{1}{24\pi^2} \lmk gn \rmk^{\frac{2}{n-2}}
                \lmk (n-2) N \rmk^{\frac{2(n-1)}{n-2}}
                v^{\frac{4(n-3)}{n-2}}, \non \\
  n_s - 1 &=& -6 \epsilon + 2 \eta \simeq 2 \eta 
          \simeq - \frac{n(n-1)g}{2^{\frac{n-4}{2}}} \frac{\phi_N^{n-2}}{v^2}
          \simeq - \frac{n-1}{n-2} \frac{2}{N}.
\eea
Inserting $\CR^2 = 2.441 \times 10^{-9}$ gives $v \simeq 6.9 \times
10^{-7}$ for $n=4, g=0.3, N=50$. In this case, there are no significant
tensor fluctuations and the spectral index becomes $n_s \simeq 0.94$,
which is on the edge of the observationally allowed region. This
situation can be improved if we take into account a higher order term $c
|\Phi|^4$ ($c = \CO(0.01)$) in the K\"ahler potential, which gives the
inflaton the additional mass slightly less than the Hubble parameter. In
this model, the gravitino mass becomes $m_{3/2} \simeq 4.3 \times
10^{-16} \simeq 1.0$~TeV.

After the inflation ends, an inflaton starts to oscillate around its
minimum $\phi_{\rm min}$ and then decays into the SM particles to reheat
the Universe. Such inflaton decay can occur, for example, by introducing
higher order terms $\sum_i c_i |\Phi|^2 |\psi_i|^2$ in the K\"ahler
potential, which are invariant under the $R$ symmetry. Here $\psi_i$
represent the SM particles and the couplings $c_i$ are of the order of
unity. Then, the decay rate of the inflaton becomes $\Gamma \simeq
\sum_i c_i^2 \phi_{\rm min}^2 m_{\phi}^3$, where $m_{\phi} \simeq n
g^{\frac{1}{n}} v^{2-\frac{2}{n}}$ is the inflaton mass at its minimum
$\phi_{\rm min}$. Thus, the reheating temperature $T_R$ is estimated as
\beq
  T_R \simeq \lmk \frac{90}{\pi^2 g_{\ast}} \rmk^{\frac14} \sqrt{\Gamma}
      \sim n^{\frac32} g^{\frac{1}{2n}} v^{3-\frac{1}{n}},
\eeq
where $g_{\ast} \simeq 200$ is the number of relativistic degrees of
freedom.  For the above parameters, $T_R \simeq 6.2 \times 10^{-17}
\simeq 1.9 \times 10^2$~GeV.

Finally, we would like to comment on the initial value problem of new
inflation. In order to have sufficiently long period of inflation, the
initial value of the inflaton must be close to the origin, that is, the
local maximum of the potential. However, since the potential is almost
flat, there is no natural reason why the inflaton initially sits close
to the origin. A couple of solutions to this initial condition problem
have been proposed so far. One of the methods is to consider another
inflation preceding new inflation \cite{Izawa:1997df,chaonew}. During
the pre-inflation, the inflaton of new inflation acquires the
aforementioned mass comparable to the Hubble parameter. This additional
mass dynamically drives it to the local minimum of the effective
potential at that time, which in turn determines initial value of new
inflation. Such a double inflation model is interesting in that it can
generate non-trivial features of primordial fluctuations, which were
studied in the context of large scale structure and primordial black
holes \cite{double}. Another method is to use the interaction with the
SM particles through the K\"ahler potential \cite{Asaka:1999yc}. Also in
this case, the inflaton of new inflation acquires the effective mass
comparable to the Hubble parameter through the interaction with the SM
particles, which are assumed to be in thermal equilibrium. This
additional mass dynamically gives adequate initial value of new
inflation again.

\subsubsection{Hybrid inflation}

\label{subsubsec:Fhybrid}

Hybrid inflation \cite{hybrid} is attractive in that it often occurs in
the context of grand unified theory (GUT) \cite{GUTinf} since its
potential can be easily associated with the spontaneous symmetry
breaking \cite{Copeland:1994vg}. Then, some other variants called
mutated hybrid \cite{Stewart:1994pt}, smooth hybrid
\cite{Lazarides:1995vr}, shifted hybrid \cite{Jeannerot:2000sv}, and
tribrid inflation \cite{Antusch:2008pn} are also proposed.

Here we consider the hybrid inflation model in supergravity proposed by
Linde and Riotto \cite{Linde:1997sj}. (See also
Refs. \cite{Panagiotakopoulos:1997qd}.) The superpotential is given by
\beq
  W =\lambda S\overline{\Psi}\Psi -\mu^2 S, 
  \label{eq:Hybridsuper}
\eeq
where $S$ is a gauge-singlet superfield, while $\Psi$ and
$\overline{\Psi}$ are a conjugate pair of superfields transforming as
nontrivial representations of some gauge group $G$. Note that, though
gauge symmetry is not always needed, this model can be easily embedded in
GUT by introducing such a gauge symmetry. $\lambda$ and $\mu$ are
positive parameters smaller than unity. This superpotential is linear in
the inflaton $S$ and possesses the $R$ symmetry, under which they are
transformed as
\beq
  S(\theta) \longrightarrow e^{2i\alpha} S(\theta e^{i\alpha}), ~~~~  
  \Psi(\theta) \longrightarrow e^{2i\alpha}\Psi(\theta e^{i\alpha}), ~~~~ 
  \overline{\Psi}(\theta) \longrightarrow 
  e^{-2i\alpha}\overline{\Psi}(\theta e^{i\alpha}).  
\eeq
Taking the canonical ($R$-invariant) K\"ahler potential,
\beq
  K = |S|^2+|\Psi|^2+|\overline{\Psi}|^2, 
  \label{eq:Hkahler}
\eeq
the scalar potential is given by the standard formulae
(\ref{eq:Fpotential}) and (\ref{eq:Dpotential}),
\bea
  V(S,\Psi,\overline{\Psi}) &=& e^{|S|^2+|\Psi|^2+|\overline{\Psi}|^2}
          \lhk (1-|S|^2+|S|^4)
          |\lambda\overline{\Psi}\Psi-\mu^2 |^2 \right. \non \\
     &&  \left.+|S|^2\lkk\left|\lambda(1+|\Psi|^2)\overline{\Psi}
         -\mu^2\Psi^\ast \right|^2
         +\left|\lambda(1+|\overline{\Psi}|^2)\Psi-\mu^2\overline{\Psi}^\ast
          \right|^2 \rkk\rhk + V_D.
 \label{eq:Hpotential}
\eea
Since the above potential does not depend on the phase of the complex
scalar field $S$, we identify its real part $\sigma \equiv \sqrt{2} {\rm
Re} S$ with the inflaton without loss of generality. Assuming $\sigma
\ll 1$, the mass matrix of $\Psi$ and $\overline{\Psi}$ is given by
\bea
  V_{\rm mass} &\simeq&
    - \lambda \mu^2 (\overline{\Psi}\Psi+\overline{\Psi}^{\ast}\Psi^{\ast})
    + \frac12 \lmk \lambda^2+\mu^4 \rmk \sigma^2
          \lmk \left|\overline{\Psi}\right|^2 +
                             \left| \Psi \right|^2  \rmk \non \\
   &=& \lkk \lmk \lambda^2+\mu^4 \rmk |S|^2 + \lambda \mu^2
       \rkk \left| \Phi \right|^2 
    +  \lkk \lmk \lambda^2+\mu^4 \rmk |S|^2 - \lambda \mu^2
       \rkk \left| \overline{\Phi} \right|^2, 
\eea
where $\Phi \equiv (\Psi - \overline{\Psi}^{\ast})/\sqrt{2}$ and
$\overline{\Phi} \equiv (\Psi + \overline{\Psi}^{\ast})/\sqrt{2}$. Thus,
we find that the eigenvalues for the corresponding eigenstates are given
by
\beq
M^2_\pm=(\lambda^2+\mu^4)|S|^2\pm\lambda\mu^2
=\frac12 (\lambda^2+\mu^4) \sigma^2\pm\lambda\mu^2,
~~~{\rm for}~~~\Psi = \mp \overline{\Psi}^{\ast},
\eeq
Since $M_{+}^2$ is always positive, we can safely set $\Phi = 0$, that
is, $\Psi = \overline{\Psi}^{\ast}$. Therefore, we have only to consider
the dynamics of $S$ and $\overline{\Phi}$.  When $\sigma$ becomes
smaller than the critical value $\sigma_c\equiv
\sqrt{2}\mu/\sqrt{\lambda}$, $M_{-}^2$ becomes negative and hence
$\overline{\Phi}$ quickly rolls down to the global minimum, which ends
inflation. This feature is typical of hybrid inflation. Here and
hereafter, we assume $\mu \ll \lambda$ for simplicity since the
extention to the other cases is straightforward.

Since $\overline{\Phi}$ is the D-flat direction, the effective potential
under the condition $\Psi=\overline{\Psi}^{\ast}$ becomes
\beq
 V = (\lambda|\Psi|^2-\mu^2)^2+\lambda^2\sigma^2|\Psi|^2+
 \frac{1}{8}\mu^4\sigma^4+\cdots.
\eeq
For $\sigma > \sigma_c$, the potential is minimized at
$\Psi=\overline{\Psi}=0$ and inflation is driven by the false vacuum
energy density $\mu^4$.

As a result of the positive energy density due to the SUSY breaking
during inflation, the mass split is induced between the scalar fields
composed of $\Psi$ and $\overline{\Psi}$ with mass squared $M^2_\pm
(\simeq \lambda^2\sigma^2/2 \pm \lambda\mu^2)$ and their superpartner
fermions with mass $M=\lambda\sigma/\sqrt{2}$ because these scalar
fields different from the inflaton receive the Hubble induced
masses. Such a mass split generates the radiative correction to the
potential. By the standard formula \cite{Coleman:1973jx}, the one-loop
correction is given by\footnote{Strictly speaking, this formula is
derived in the Minkowski background. Therefore, we need to modify it for
the De Sitter background, where inflation happens.}
\beq
 V_{1L}=\frac{\lambda^2 \CN}{128\pi^2}\lkk (\lambda\sigma^2+2\mu^2)^2
 \ln\frac{\lambda\sigma^2+2\mu^2}{\Lambda^2}
+(\lambda\sigma^2-2\mu^2)^2
 \ln\frac{\lambda\sigma^2-2\mu^2}{\Lambda^2}
-2\lambda^2\sigma^4\ln\frac{\lambda\sigma^2}{\Lambda^2}\rkk,
  \label{eq:oneloop}
\eeq
where $\Lambda$ is some renormalization scale and $\CN$ stands for the
dimensionality of the representation of the gauge group $G$ to which the
fields $\Psi$ and $\overline{\Psi}$ belong. When $\sigma \gg \sigma_c$,
it is approximated as
\beq
 V_{1L} \simeq \frac{\lambda^2 \CN \mu^4}{8\pi^2}\ln\frac{\sigma}{\sigma_c}.
 \label{eq:Lapprox}
\eeq
Thus, the effective potential of the inflaton $\sigma$ during inflation
is given by
\beq
 V(\sigma) \simeq \mu^4\lmk 1+\frac{\wlambda^2}{8\pi^2}
                  \ln\frac{\sigma}{\sigma_c} +\frac{1}{8}\sigma^4 \rmk,
\label{eq:Hpotea}
\eeq
with $\wlambda \equiv \lambda \sqrt{\CN}$. This effective potential is
dominated by the false vacuum energy $\mu^4$ for $\sigma \ll 1$. On the
other hand, the dynamics of $\sigma$ is determined by the competition
between the second and the third terms of the right hand side. Comparing
their first derivatives, we find that the dynamics of the inflaton
$\sigma$ is dominated by the radiative correction for $\sigma <
\sqrt{\wlambda/(2\pi)}\equiv\sigma_d$ and by the non-renormalizable term
for $\sigma > \sigma_d$. Notice that $\sigma_c \ll \sigma_d \ll 1$ for
$\mu \ll \wlambda \ll 1$. Then, the total number of $e$-folds during
inflation, $N_{\rm total}$, is given by
\beq
 N_{\rm total} = \int_{\sigma_c}^{\sigma_i}\frac{V}{V'} d\sigma
 \simeq \int_{\sigma_c}^{\sigma_d}\frac{8\pi^2}{\wlambda^2}\sigma d\sigma
       + \int_{\sigma_d}^{\sigma_i}\frac{2}{\sigma^3}d\sigma
 \simeq \frac{2\pi}{\wlambda} +\frac{2\pi}{\wlambda} =  \frac{4\pi}{\wlambda}, 
 \label{eq:Hefold}
\eeq 
where $\sigma_i$ is the initial value of hybrid inflation. We find that
the amount of inflation during $\sigma>\sigma_d$ and that during
$\sigma<\sigma_d$ are about the same with the $e$-folding number $\simeq
2\pi/\wlambda$.  Thus in order to achieve sufficiently long inflation
$N_{\rm total} \gtrsim 60$ to solve the horizon and flatness problems,
$\wlambda$ should be rather small: $\wlambda \lesssim 4\pi/60 \simeq
0.2$. In particular, in case that $\wlambda \lesssim 2\pi/60 \simeq
0.1$, the dynamics of the inflaton $\sigma$ for the observable Universe
is determined only by the radiative correction. First of all, we
concentrate on this case. Then, the $e$-folding number is estimated as
\beq
 N = \int_{\sigma_c}^{\sigma_N}\frac{V}{V'} d\sigma
 \simeq \int_{\sigma_c}^{\sigma_N}\frac{8\pi^2}{\wlambda^2}\sigma d\sigma
 \simeq \frac{4\pi^2}{\wlambda^2} \sigma_N^2,
\eeq
which yields $\sigma_N \simeq \wlambda \sqrt{N}/(2 \pi)$. Here
$\sigma_N$ is a field value corresponding to the $e$-folding number
equal to $N$. The slow-roll parameters are estimated as
\beq
  \epsilon \simeq \frac{\wlambda^4}{128\pi^4 \sigma_N^2}
           \simeq \frac{\wlambda^2}{32\pi^2 N},~~~~
  \eta \simeq - \frac{\wlambda^2}{8\pi^2\sigma_N^2} \simeq - \frac{1}{2N}.
\eeq 
The amplitude and the spectral index of primordial fluctuations are
given by
\bea
  \CR^2 &=& \frac{1}{24 \pi^2} \frac{V}{\epsilon}
        \simeq \frac{16 \pi^2}{3} \frac{\mu^4 \sigma_N^2}{\wlambda^4}
        \simeq \frac{4\mu^4}{3\wlambda^2} N, \non \\
  n_s - 1 &=& -6 \epsilon + 2 \eta \simeq 2 \eta \simeq - \frac{1}{N},
           \non \\
  r &=& 16 \epsilon = \frac{\wlambda^2}{2\pi^2 N}. 
\eea
Taking into account $\CR^2 = 2.441 \times 10^{-9}$, we obtain $\mu
\simeq 2.4 \times 10^{-3} \sqrt{\wlambda}$ for $N = 60$ and have
negligible tensor perturbations. In addition, the spectral index becomes
$n_s \simeq 0.98$ for $N \simeq 60$ and is just outside the
observationally allowed values. However, numerical calculations by using
the full one-loop potential (\ref{eq:oneloop}) are necessary for more
accurate estimates because the approximate form of the one-loop
potential (\ref{eq:Lapprox}) is justified only for $\sigma \gg
\sigma_c$. Such detailed calculations gave the similar results $n_s
\gtrsim 0.985$. Therefore, more elaborated models to obtain a lower
spectral index were proposed in Refs. \cite{Jeannerot:2005mc}.

In fact, we also need to take into account the formation of topological
defects at the end of inflation, which is associated with the gauge
symmetry breaking. It was pointed out that, if we try to embed this
model of hybrid inflation into a realistic model of supersymmetric GUT,
the formation of cosmic strings is inevitable
\cite{Jeannerot:2003qv,Rocher:2004et}. Such cosmic strings can
contribute to the CMB anisotropies, which gives a severe constraint on
the model parameters. Although the prediction of the contribution of
cosmic strings to the CMB anisotropies still have some uncertainties, it
cannot exceed $\sim 10$\% \cite{Bouchet:2000hd,Endo:2003fr}, which gives
the constraint on the coupling constant $\lambda$ \cite{Rocher:2004et},
\beq
  \lambda \lesssim 7 \times 10^{-7} \times \frac{126}{\CN}.
\eeq
For the $SO(10)$ GUT model with $\CN = 126$, $\lambda \lesssim 7 \times
10^{-7}$.

In the case that $\wlambda \gtrsim 2\pi/60 \simeq 0.1$, we need to
consider both contributions coming from the radiative correction and the
non-renormalizable term. Then, the $e$-folding number is estimated as
\beq
 N = \int_{\sigma_c}^{\sigma_N}\frac{V}{V'} d\sigma
 \simeq \int_{\sigma_c}^{\sigma_d}\frac{8\pi^2}{\wlambda^2}\sigma d\sigma
       + \int_{\sigma_d}^{\sigma_N}\frac{2}{\sigma^3}d\sigma
 \simeq \frac{4\pi}{\wlambda} - \frac{1}{\sigma_N^2},
\eeq
where $\sigma_N$ is a field value corresponding to the $e$-folding
number equal to $N$. Calculating the slow-roll parameters $\epsilon$ and
$\eta$ for the potential (\ref{eq:Hpotea}), we find
\beq
 \epsilon = \frac{1}{8\sigma^2} \lmk \sigma_d^4+\sigma^4 \rmk^2,~~~
 \eta = \frac{1}{2\sigma^2} \lmk -\sigma_d^4+3\sigma^4 \rmk^2.
\eeq
The amplitude of primordial density fluctuations becomes
\beq
  \CR^2 = \frac{1}{24\pi^2} \frac{V}{\epsilon} 
        = \frac{\mu^4}{3\pi^2} 
          \frac{\sigma^2}{\lmk \sigma_d^4+\sigma^4 \rmk^2}.
\eeq
Inserting $\CR^2 \simeq 2.441 \times 10^{-9}$ yields $\mu \simeq 1.4
\times 10^{-3} \simeq 3.3 \times 10^{15}$~GeV and $\sigma_N \simeq 0.17$
for $N = 60$ and $\wlambda =0.13$. The total $e$-folding number is
$N_{\rm total} \simeq 97$, though more correct values need numerical
calculations. On the other hand, the spectral index of scalar
perturbation is given by
\bea
 n_s-1 = - 6 \epsilon + 2 \eta
       \simeq 2 \eta 
       = 3 \sigma^2 - \frac{\sigma_d^4}{\sigma^2}. 
\label{eq:Hns}
\eea
Interestingly, the spectral index crosses unity at
$\sigma=\sigma_d/3^{1/4}\sim 0.8\sigma_d$. This is mainly because the
spectral index is smaller than unity in the region where the radiative
correction dominates, while it is larger than unity in the region where
the non-renormalizable term dominates. Such feature also suggests that
this model can generate the large running of the spectral index. Such a
large running of the spectral index was first suggested by the WMAP
first year result \cite{Peiris:2003ff} and is slightly preferred even by
the WMAP seven year result with the ACT 2008 data \cite{Dunkley:2010ge},
which yields $n_s = 1.032 \pm 0.039$ and $d n_s/d \ln k =
-0.034\pm0.018$ at the pivot scale $k_0 = 0.002 {\rm Mpc}^{-1}$ with $68
\%$ CL. The running of the spectral index can be evaluated by use of the
slow-roll parameters,
\beq
  \frac{dn_s}{d\ln k} = 16 \epsilon \eta - 24 \epsilon^2 - 2\xi,~~~~~
  \xi \equiv \frac{V''' V'}{V^2}.
\eeq
For $\wlambda = 0.13$, $d n_s/d \ln k \simeq -4.2 \times 10^{-3}$, whose
magnitude is too small to explain the suggested running. As $\wlambda$
becomes larger, the running gets bigger. But, large $\wlambda$ leads to
a small number of the total $e$-folds $N_{\rm total}$. In fact, it is
found that $N_{\rm total}$ is at most $20$ in order to accommodate the
running with $d n_s/d \ln k = \CO(10^{-2})$, which needs another
inflation after hybrid inflation to solve the flatness and horizon
problems \cite{Kawasaki:2003zv}. See Refs. \cite{Feng:2003mk} for other
attempts to explain such large running.

After the inflation ends, the inflaton and the field $\overline{\Phi}$
start to oscillate around their minima and then decay into the SM
particles to reheat the Universe. Since the field $\overline{\Phi}$
acquires a non-zero vacuum expectation value (VEV), such decay happens
by introducing the higher order terms $\sum_i c_i |\overline{\Phi}|^2
|\psi_i|^2$ in the K\"ahler potential, which are invariant under the $R$
symmetry. Here $\psi_i$ represent the SM particles and the couplings
$c_i$ are of the order of unity. The decay rate of the inflaton becomes
$\Gamma \simeq \sum_i c_i^2 \overline{\Phi}_{\rm min}^2
m_{\overline{\Phi}}^3$, where $\overline{\Phi}_{\rm min} \simeq \sqrt{2}
\mu/\sqrt{\wlambda}$ and $m_{\overline{\Phi}} \simeq \sqrt{2 \wlambda}
\mu$. Then, the reheating temperature $T_R$ is evaluated as
\beq
  T_R \simeq \lmk \frac{90}{\pi^2 g_{\ast}} \rmk^{\frac14} \sqrt{\Gamma}
      \sim C \wlambda^{\frac14} \mu^{\frac52},
\eeq
with $C \equiv \sqrt{\sum_i c_i^2}$.  For the above parameters, $T_R
\simeq 4.2~C \times 10^{-8} \simeq 1.0~C \times 10^{11}$~GeV, which is
relatively high. \\

In the above model of hybrid inflation, quantum correction such as the
one-loop correction plays an important role in the dynamics. Then, you
may wonder whether quantum correction could make effective potential
flat even if classical potential is steep. Such possibility is actually
what was pointed out in Ref. \cite{Stewart:1996ey} and is called running
mass inflation. This is another solution to the $\eta$ problem. Let us
consider a potential of hybrid inflation type,
\beq
  V(\sigma,\psi) = V_0 f(\psi) + \frac12 m^2 \sigma^2 
                   + \frac12 g^2 \sigma^2 \psi^2,
\eeq
where $\sigma$ is an inflaton and $\psi$ is a waterfall field staying
at the origin during inflation. $f(\psi)$ is a function of $\psi$ and
takes a maximum value $f(\psi=0) = 1$ at $\psi = 0$. Below some critical
value $\sigma_c$, $\psi$ is destabilized and quickly rolls down to the
global minimum of $f(\psi)$. $g$ is a coupling constant. $m$ is a tree
level mass of the inflaton $\sigma$ and could be as large as the Hubble
parameter, which would rule out $\sigma$ as an inflaton. However,
suppose that the SUSY be explicitly (softly) broken during inflation,
quantum correction generates an additional mass squared,
\beq
 m^2(\sigma) = m^2 
   + \frac{\lambda}{32\pi^2} \widetilde{m}^2 \ln \frac{\sigma}{\Lambda}.
\eeq
Here $\widetilde{m}^2$ is the soft mass squared, $\lambda$ is a Yukawa
or gauge coupling constant, which is not too small, and $\Lambda$ is the
renormalization scale. Then, the potential near $m^2(\sigma) = 0$ is
flat enough to support inflation even if the tree level mass $m$ is too
large. This is the essential idea of the running mass inflation. Making
adequate changes of the variables and parameters, the effective
potential of the inflaton $\sigma$ can be recast into \cite{Covi:2004tp}
\beq
  V(\sigma) = V_0 + \frac12 m^2(\sigma) \sigma^2 
            = V_0 \lkk 1 + \frac12 \eta_0 \sigma^2 
               \lmk \ln \frac{\sigma}{\sigma_{\ast}} - \frac12 \rmk
  \rkk.
\eeq
We can easily find that $m^2(\sigma_0)=0$ at $\ln \sigma_0/\sigma_{\ast}
= 1/2$ and
\beq
  \frac{V'}{V_0} = \eta_0 \sigma \ln \frac{\sigma}{\sigma_{\ast}},
\eeq
which implies that the potential takes a maximum value at $\sigma =
\sigma_{\ast}$ by assuming $\eta_0 < 0$. The slow-roll parameters are
given by
\bea
  \epsilon &\simeq& \frac12
           \lmk \eta_0 \sigma \ln \frac{\sigma}{\sigma_{\ast}} \rmk^2
           = \CO \lmk \eta_0^2 \sigma_{\ast}^2 \rmk, \non \\
  \eta &\simeq& \eta_0 \lmk 1 + \ln \frac{\sigma}{\sigma_{\ast}} \rmk 
           = \CO(\eta_0), \non \\  
  \xi &\simeq& \eta_0^2 \ln \frac{\sigma}{\sigma_{\ast}}
           = \CO(\eta_0^2).
\eea 
Then, the spectral index and its running are calculated as
\bea
  n_s - 1 &=& - 6\epsilon + 2\eta \simeq 2 \eta 
      \simeq 2 \eta_0 \lmk 1 + \ln \frac{\sigma}{\sigma_{\ast}} \rmk, \non \\
  \frac{dn_s}{d\ln k} &=& 16 \epsilon \eta - 24 \epsilon^2 - 2\xi \simeq -2 \xi
      \simeq -2 \eta_0^2 \ln \frac{\sigma}{\sigma_{\ast}}.
\eea
From the last equation, we find that the negative running is obtained
for the region with $\sigma > \sigma_{\ast}$. Taking $|\eta_0| =
\CO(0.01)$ gives $n_s-1 = \CO(0.01)$ with the negligible running, which
is compatible with the observational results. On the other hand, if we
take $|\eta_0| = \CO(0.1)$, $n_s-1 = \CO(0.1)$ and $d n_s/d \ln k =
\CO(10^{-2})$ as suggested in \cite{Dunkley:2010ge}. The $e$-folding
number is given by
\beq
  N \simeq \int_{\sigma_e}^{\sigma_N} \frac{V}{V'} d\sigma
    = \frac{1}{\eta_0} \int_{\sigma_e}^{\sigma_N} 
        \frac{d\sigma}{\sigma \ln \frac{\sigma_{\ast}}{\sigma}}
   = - \frac{1}{|\eta_0|} \lkk \ln \lmk
         \frac{ \ln \frac{\sigma_N}{\sigma_{\ast}} }
              { \ln \frac{\sigma_e}{\sigma_{\ast}} }
                                   \rmk \rkk.
\label{eq:Nrun}
\eeq
Here $\sigma_e$ is the field value of $\sigma$ at the end of inflation,
determined either by the critical value $\sigma_c$ or by the violation
of the slow-roll conditions. Then, the following condition must be
satisfied at least,
\beq
  |\eta_e| 
   = \left| \eta_0 \right| \lmk 1 + \ln \frac{\sigma_e}{\sigma_{\ast}} \rmk
   \simeq \left| \eta_0 \right| \ln \frac{\sigma_e}{\sigma_{\ast}} 
   \lesssim 1.
\eeq
Inserting Eq. (\ref{eq:Nrun}) into this condition yields
\beq
  \ln \frac{\sigma_N}{\sigma_{\ast}} 
   = e^{-N |\eta_0|} \ln \frac{\sigma_e}{\sigma_{\ast}} 
   \lesssim \frac{1}{|\eta_0|} e^{-N |\eta_0|}
   \lesssim 1,~~~~~{\rm for}~~~|\eta_0| \sim 0.1,
\eeq
which shows that $\sigma_N \sim \sigma_{\ast}$ for $|\eta_0| \sim
0.1$. The amplitude of the primordial density fluctuations becomes
\beq
  \CR^2 \simeq \frac{1}{24 \pi^2} 
      \frac{V_0}{\lmk \eta_0 \sigma_N \ln \frac{\sigma_N}{\sigma_{\ast}} \rmk^2}.
\eeq
Inserting $\CR^2 \simeq 2.441 \times 10^{-9}$ gives the relation
$V_0^{1/4} \simeq 2.8 \times 10^{-2} \sqrt{ \eta_0 \sigma_N \ln
\frac{\sigma_N}{\sigma_{\ast}} }$. For $|\eta_0| \sim 0.1$ and $\sigma_N
\sim \sigma_{\ast}$, $V_0^{1/4} \sim 8.7 \times 10^{-3} \sigma_{\ast}^{1/2}$.

Like new inflation, hybrid inflation may have severe initial value
problem, which states that only very narrow range of initial field
values (less than unity) can lead to successful inflation
\cite{Lazarides:1997vv}. Though the recent paper claims the opposite
result \cite{Clesse:2009ur}, there is still another initial value
problem, that is, why an inflaton is homogeneous over the Hubble horizon
scale at the onset of inflation. The initial value problem may be solved
by considering pre-inflation preceding hybrid inflation
\cite{Panagiotakopoulos:1997if} in the same way as new inflation.  Such
double inflation scenario can produce non-trivial features of primordial
fluctuations such as the break of the spectral index and the formation
of the primordial black holes.

\subsubsection{Chaotic inflation}

\label{subsubsec:Fchaotic}

Chaotic inflation is the most natural inflation in that it does not
suffer from any initial condition problem simply because it can start
around the Planck time or the Planck energy density scale. Since the
other inflations occur at later time or lower scale, the Universe would
recollapse before inflation starts, unless the Universe is open at the
beginning. In addition, the other models suffer from the initial value
problem, that is, why the inflaton field is homogeneous over the horizon
scale and takes a value which can lead to successful inflation.

A simple power-law potential $V(\phi) = \lambda_n \phi^n / n$
($\lambda_n \ll 1$ : a coupling constant) can accommodate chaotic
inflation. The slow-roll parameters are given by
\beq
  \epsilon \simeq \frac{n^2}{2 \phi^2},~~~~
  \eta \simeq \frac{n(n-1)}{\phi^2},
\eeq
which requires the field value of the inflaton to be larger than unity
for successful inflation. That is, chaotic inflation started around the
Planck scale, where the field value is much larger than unity (and also
is stochastic), and then it ended around $\phi_e \sim n$. The
$e$-folding number becomes $N \simeq \phi_N^2 / (2n)$ and the observable
quantities are expressed in terms of $N$ as
\bea
  \CR^2 &=& \frac{1}{24\pi^2}\frac{V}{\epsilon} 
        = \frac{\lambda_n}{12\pi^2 n^3} \phi_N^{n+2}
        \simeq \frac{\lambda_n}{12\pi^2 n^3} (2nN)^{n+2}, \non \\ 
  n_s - 1 &=& -6\epsilon+2\eta = -\frac{n(n+2)}{\phi_N^2} 
          \simeq - \frac{n+2}{2N}, \non \\
  r &=& 16 \epsilon = \frac{8n^2}{\phi_N^2} \simeq \frac{4n}{N}.
  \label{eq:chaoticobs}
\eea
The constraint on the tensor-to-scalar ratio $r < 0.24$ leads to $n <
0.06 N \lesssim 3.6$ for $N \lesssim 60$, which rules out the case with
$n \ge 4$ if the primordial density fluctuations are mainly generated by
the inflaton. Though another source of the primordial density
fluctuations like the curvaton and the modulated reheating mechanism can
save the case with $n \ge 4$ \cite{Langlois:2004nn}, we do not consider
such a case in this review.

As shown above, a large field value of a would-be inflaton is required
to cause chaotic inflation. However, the exponential factor appearing in
the F-term potential prevents any scalar field from taking a value
larger than unity, provided that the K\"ahler potential is almost
canonical. Thus, it is extremely difficult to incorporate chaotic
inflation in supergravity, even if we can solve the $\eta$ problem
somehow. Though chaotic inflation was proposed in some models
\cite{Goncharov:1983mw}, most of them used rather specific K\"ahler
potential, which was fine-tuned without the symmetry reason.

In this review, following Refs. \cite{Kawasaki:2000yn}, we introduce the
Nambu-Goldstone-like shift symmetry of the inflaton superfield $\Phi$ in
order to naturally realize chaotic inflation in supergravity. We require
the K\"ahler potential $K(\Phi,\Phi^{\ast})$ to be invariant under the
shift of $\Phi$,
\beq
  \Phi \longrightarrow \Phi + i~C,
  \label{eq:shift}
\eeq
where $C$ is a real parameter. Then, the K\"ahler potential is a
function of $\Phi + \Phi^{\ast}$, that is, $K(\Phi,\Phi^{\ast}) = K(\Phi
+ \Phi^{\ast})$, which implies that the imaginary part of the scalar
components of $\Phi$ does not appear in the exponential factor of the
F-term and hence can take a value larger than unity. Notice that this
model solves the $\eta$ problem as well. However, as long as the shift
symmetry is exact, the potential is completely flat along the inflaton
direction. Therefore, in order to cause inflation, a small breaking term
of the shift symmetry must be introduced. As such a breaking term, we
consider a small mass term in the superpotential by introducing another
superfield $X$,
\beq
  W = mX\Phi.
  \label{eq:chaoticmass}
\eeq
Note that this model is natural in 't Hooft's sense \cite{tHooft}
because we have an enhanced symmetry (the shift symmetry) in the limit
$m \rightarrow 0$. Neglecting the induced breaking terms such as $K
\simeq |m\Phi|^{2}+\cdots$, which are irrelevant for the dynamics of the
inflaton, we consider the following K\"ahler potential,
\beq
  K = \frac12 (\Phi + \Phi^{\ast})^{2} + XX^{\ast},
  \label{eq:chaotickahler}
\eeq
which gives the canonical kinetic terms for $\Phi$ and $X$. This model
possesses the $R$ symmetry under which
\beq
    X(\theta) \longrightarrow e^{2i\alpha} X(\theta e^{i\alpha}),~~~~
    \Phi(\theta) \longrightarrow \Phi(\theta e^{i\alpha}),
  \label{eq:Rsymmetry}
\eeq
and $Z_{2}$ symmetry under which
\beq
    X(\theta) \longrightarrow - X(\theta),~~~~ 
    \Phi(\theta) \longrightarrow - \Phi(\theta).
  \label{eq:Z2symmetry}
\eeq
By inserting the forms of the K\"ahler potential and the superpotential
into the formulae (\ref{eq:kinetic}) and (\ref{eq:Fpotential}), the
Lagrangian density $L(\varsigma,\varphi,X)$ becomes
\begin{equation}
  L(\varsigma,\varphi,X) = -\frac12 \partial_{\mu}\varsigma\partial^{\mu}\varsigma 
              - \frac12 \partial_{\mu}\varphi\partial^{\mu}\varphi 
              - \partial_{\mu}X\partial^{\mu}X^{*}
              -V(\varsigma,\varphi,X),
  \label{eq:Lagrangian2}
\end{equation}
with the potential $V(\varsigma,\varphi,X)$ given by
\begin{eqnarray}
  \lefteqn{V(\varsigma,\varphi,X) = m^{2} \exp \left( \varsigma^{2} + |X|^{2}
                                           \right) } \nonumber \\ 
     &&    \times \left[ ~ 
                 \frac{1}{2} (\varsigma^{2}+\varphi^{2})(1+|X|^{4}) 
               + |X|^{2} \left\{
                 1 - \frac12(\varsigma^{2}+\varphi^{2})
               + 2 \varsigma^{2} \left( 1 + \frac12(\varsigma^{2}+\varphi^{2}) \right)
                         \right\} ~\right].
  \label{eq:potential2}
\end{eqnarray}     
Here we have decomposed $\Phi$ into a real part $\varsigma$ and an
imaginary one $\varphi$,
\beq
  \Phi = \frac{1}{\sqrt{2}} (\varsigma + i \varphi),
\eeq
and identify $\varphi$ with the inflaton. While the inflaton $\varphi$
can have a value much larger than unity, $|\varsigma|, |X| \lesssim 1$
because of the presence of $e^{K}$ factor. Then, the potential is
rewritten as
\beq
  V(\varsigma,\varphi,X) \simeq \frac12 m^2 \varphi^2 (1+\varsigma^2) + m^2 |X|^2.
\eeq
The effective mass of $\varsigma$ is larger than the Hubble parameter during
inflation, it quickly rolls down to zero. On the other hand, though the
effective mass of $X$ is not larger than the Hubble
parameter,\footnote{If we take into account a higher order term $- c
|X|^4$ ( $c \gtrsim 1$: positive constant) in the K\"ahler potential, it
gives $X$ an additional mass larger than the Hubble parameter, which
quickly drives $X$ to zero.} we can easily show that it is irrelevant
for the dynamics and the primordial density fluctuation. Then, $X$ can
be safely set to zero, which yields, together with $\varsigma = 0$,
\beq
  V(\varphi) \simeq \frac 12 m^2 \varphi^2.
\eeq
Thus, a simple power-law potential can be naturally obtained in the
context of supergravity. 

After the inflation ends, an inflaton starts to oscillate around the
origin and then decays into the SM particles to reheat the
Universe. Such inflaton decay can occur, for example, by introducing the
following superpotential,
\beq
    W = g X H_u H_d.
\eeq
Here $H_u$ and $H_d$ are a pair of Higgs doublets, whose $R$ charges
are assumed to be zeroes, and $g$ is a coupling constant. Then, we have
a coupling of the inflaton $\varphi$ to the Higgs doublets as
\beq
    L \sim  g m \varphi H_u H_d,
\eeq
which gives the reheating temperature 
\beq
    T_R \sim 10^{9}~{\rm GeV} 
    \left(\frac{g}{10^{-5}}\right)
    \left(\frac{m}{10^{13}{\rm GeV}}\right)^{1/2}.
\eeq
By inserting $n=2$ and $N = 60$ to the formulae (\ref{eq:chaoticobs}),
we obtain the spectral index $n_s \simeq 0.967$ and the inflaton mass $m
\equiv \sqrt{\lambda_2} \simeq 6.3 \times 10^{-6} = 1.5 \times
10^{13}$~GeV. The large tensor-to-scalar ratio is also predicted to be
$r \simeq 0.13$, which is still compatible with the present observations
and will be detected or ruled out. Then, it may be interesting to
consider chaotic inflation model with a lower $n$ and even a fractional
one. In fact, such a chaotic inflation model with a lower power-law
index was proposed in the context of superstring
\cite{Silverstein:2008sg}. \\

Even in the context of supergravity, it is easy to realize such a model
of chaotic inflation by extending the Nambu-Goldstone-like shift
symmetry, which is named running kinetic inflation. According to
Ref. \cite{Takahashi:2010ky}, we impose the following type of shift
symmetry on a composite field $\Phi^2$,
\beq
  \Phi^2 \longrightarrow \Phi^2 + C,
  \label{eq:shift2}
\eeq
where $\Phi \ne 0$ and $C$ is a real parameter. Then, the K\"ahler
potential is a function of $\lmk \Phi^2 - \Phi^{\ast 2} \rmk$, that is,
\beq
  K(\Phi,\Phi^{\ast}) = i c \lmk \Phi^2 - \Phi^{\ast 2} \rmk 
    - \frac14 \lmk \Phi^2 - \Phi^{\ast 2} \rmk^2 + \cdots,
\eeq
where $c$ is a real parameter of the order of unity. As long as this
shift symmetry is exact, the potential is completely flat along the
direction of the shift and the kinetic term is singular at the
origin. Then, small breaking terms of the shift symmetry are necessary
for the K\"ahler potential as well as the superpotential. We add the
following breaking term to the K\"ahler potential,
\beq
  K = \kappa |\Phi|^2,
\eeq
with $\kappa \ll 1$. This term cures the singular behavior of the
kinetic term of $\Phi$ at the origin. On the other hand, we consider a
small mass term as a breaking term in the superpotential,
\beq
  W = mX\Phi.
  \label{eq:chaoticmass2}
\eeq
Here $m \ll 1$ and we have introduced another superfield $X$, which is
assumed to have the canonical K\"ahler potential. Notice that, though we
have introduced the breaking terms, this model is still natural in 't
Hooft's sense \cite{tHooft} because we have an enhanced symmetry (the
shift symmetry) in the limit $m \rightarrow 0$ and $\kappa \rightarrow
0$. Then, the total K\"ahler potential we consider here is given by
\beq
  K = \kappa |\Phi|^2 + i c \lmk \Phi^2 - \Phi^{\ast 2} \rmk 
    - \frac14 \lmk \Phi^2 - \Phi^{\ast 2} \rmk^2 + |X|^2,
  \label{eq:chaotickahler2}
\eeq
which yields the following kinetic terms,
\bea
  \CL_{\rm kin} &=& 
    - \lmk \kappa+2|\Phi|^2 \rmk \del_{\mu}\Phi \del^{\mu}\Phi^{\ast}
   - \del_{\mu}X \del^{\mu}X^{\ast} \non \\
                &=& 
    - \frac12 \lmk \kappa^2 + \phi^2 + \chi^2 \rmk 
            \lmk \del_{\mu}\phi \del^{\mu}\phi 
                 + \del_{\mu}\chi \del^{\mu}\chi \rmk           
   - \del_{\mu}X \del^{\mu}X^{\ast}.
\eea
Here we have decomposed $\Phi$ into a real part $\phi$ and an imaginary
one $\chi$,
\beq
  \Phi = \frac{1}{\sqrt{2}} (\phi + i \chi),
\eeq
and identify $\phi$ with the inflaton. Though the potential term is a
bit complicated, we can safely set $X = 0$ in the same way as the
original chaotic inflation model in supergravity. Then, the potential at
$X=0$ reads
\beq
  V = e^K m^2 |\Phi|^2
    = \frac12 \exp \lkk 
        \frac{\kappa}{2} \lmk \phi^2+\chi^2 \rmk
       - 2 c \phi \chi + \phi^2 \chi^2
                   \rkk
      m^2 \lmk \phi^2 + \chi^2 \rmk.
\eeq 
You can find the flat direction $\phi \chi =$~constant, reflecting the
shift symmetry (\ref{eq:shift2}). For a large value of $|\phi|$, $\chi$
is determined such that the K\"ahler potential takes a minimum, that is,
\beq
  \chi \simeq \frac{c}{\phi}.
\eeq
In this limit $|\phi| \gg 1$, the potential reduces to
\beq 
  V \simeq 
     \exp \lkk 
     \frac{\kappa}{2} \lmk \phi^2 +\frac{c^2}{\phi^2} \rmk - c^2
          \rkk m^2 \lmk \phi^2 +\frac{c^2}{\phi^2} \rmk. 
\eeq
Since $\kappa \ll 1$, the exponential factor is at most of the order of
unity for $|\phi| \lesssim 1/\sqrt{\kappa}$. Then, for $1 \ll \phi \ll
1/\sqrt{\kappa}$, the effective Lagrangian is approximated as
\beq
  \CL \simeq - \frac12 \phi^2 \del_{\mu}\phi \del^{\mu}\phi
             - \frac12 \overline{m}^2 \phi^2,
\eeq
with $\overline{m}^2 \equiv e^{-c^2} m^2$. By taking the canonically
normalized field $\varphi \equiv \phi^2/2$, the effective Lagrangian is
rewritten as
\beq
  \CL \simeq - \frac12  \del_{\mu}\varphi \del^{\mu}\varphi 
                - \overline{m}^2 \varphi,
\eeq 
for $1 \ll \varphi \ll 1/\kappa$. Thus, chaotic inflation with
the linear potential is realized in supergravity. If we introduce
another type of shift symmetry, under which
\beq
  \Phi^n \longrightarrow \Phi^n + C,
  \label{eq:shiftn}
\eeq
and the superpotential $W \propto \Phi^m X $, we have chaotic inflation
with the effective potential $V \propto \varphi^{2m/n}$. Thus, this
model can possess even a fractional power. \\

Although it is not usually stressed, the fact is crucially important
that the superpotential depends on another field $X$ other than the
inflaton and is linear in it, as given in the previous examples. This
superpotential vanishes at $X=0$, which guarantees the positivity of the
potential. In fact, a lot of attempts to realize chaotic inflation in
supergravity have been hindered by the dangerous negative term $-3
|W|^2$ of the potential in supergravity. On the other hand, the fact
that $\del W/\del X$ does not depend on $X$ allows the inflaton to
acquire the potential depending on only the inflaton. Recently, by using
this type of superpotential, Kallosh {\it et al.} gave the prescription
to construct arbitrary potential in supergravity
\cite{Kallosh:2010xz}. According to Ref. \cite{Kallosh:2010xz}, we give
such a prescription and the criterion for the stability of the given
potential.

First of all, we consider the superpotential linear in $X$,
\beq
  W = X f(\Phi), 
\eeq
where $f(\Phi)$ is a real holomorphic function, that is, $f^{\ast}(\Phi)
= f(\Phi)$. This type of superpotential can be easily obtained by
imposing the $R$ symmetry with the $R$ charges of $X$ and $\Phi$ to be
$2$ and $0$, respectively. The real part of $\Phi$ will be identified
with the inflaton. On the other hand, Im $\Phi$ and $X$ will be set to
zeroes during inflation.

We assume that the K\"ahler potential $K(\Phi,\Phi^{\ast},X,X^{\ast})$
is separately invariant under the following transformations,
\beq
 X \longrightarrow -X,~~~
 \Phi \longrightarrow \Phi^{\ast},~~~ 
 \Phi \longrightarrow \Phi + C, 
\eeq
where $C$ is a real parameter. This $Z_2$ symmetry on $X$ guarantees
$K_X=K_{X^{\ast}}=K_{X\Phi^{\ast}}=K_{X^{\ast}\Phi}=0$ and $\del V/\del
X=\del V/\del X^{\ast} = 0$ at $X=0$. Since $D_{\Phi} W = 0$ and $D_{X}
W = f(\Phi)$ at $X=0$, the potential at $X=0$ becomes
\beq
  V = e^{K(\Phi,\Phi^{\ast},0,0)} f^2(\Phi)
       K_{XX^{\ast}}^{-1}(\Phi,\Phi^{\ast},0,0). 
\eeq
By virtue of the shift symmetry, the K\"ahler potential only depends on
Im $\Phi$ at $X=0$. The effective masses squared of Im $\Phi$ and $X$
become positive and larger than the Hubble parameter squared under some
conditions, which will be discussed later. In this case, we can safely
set Im $\Phi$ and $X$ to be zeroes. Then, the potential at Im $\Phi=X=0$
is rewritten as
\beq
  V = e^{K(0,0,0,0)} f^2({\rm Re}~\Phi) K_{XX^{\ast}}^{-1}(0,0,0,0). 
\eeq
By use of the K\"ahler transformation, we can always set $K_{(0,0,0,0)}
= 0$, corresponding to the rescaling of $f(\Phi)$. By rescaling of the
fields, we can also set $K_{XX^{\ast}}^{-1}(0,0,0,0) =
K_{\Phi\Phi^{\ast}}^{-1}(0,0,0,0) = 1$, which, together with
$K_{X\Phi^{\ast}}=K_{X^{\ast}\Phi}=0$ at $X=0$, means the canonical
kinetic terms of $X$ and $\Phi$ at the origin. Thus, the potential
reduces to
\beq
  V = f^2({\rm Re}~\Phi) \ge 0.
\eeq
By decomposing the complex scalar field $\Phi$ into a real part $\phi$
and an imaginary one $\chi$,
\beq
  \Phi = \frac{1}{\sqrt{2}} \lmk \phi + i \chi \rmk,
\eeq
the potential is rewritten in terms of the inflaton $\phi$ as
\beq
  V = f^2 \lmk \frac{\phi}{\sqrt{2}} \rmk \ge 0.
\eeq
In order to investigate the stability conditions during inflation, after
some calculations, we obtain the effective masses squared at $\chi=X=0$
(namely the inflationary trajectory Im $\Phi=X=0$),
\bea
 && m_{\chi}^2 = 2 \lmk 1-K_{\Phi\Phi^{\ast}XX^{\ast}} \rmk f^2
                + \lmk \frac{df}{d\Phi} \rmk^2 - f \frac{d^2f}{d\Phi^2}
                \simeq 3 H^2 \lkk 2 \lmk 1-K_{\Phi\Phi^{\ast}XX^{\ast}} \rmk
                                  2 \epsilon - \eta \rkk,   \non \\
 && m_{X}^2 = 
   - K_{XXX^{\ast}X^{\ast}} f^2 + \lmk \frac{df}{d\Phi} \rmk^2
   \simeq 3 H^2 \lmk - K_{XXX^{\ast}X^{\ast}} +\epsilon \rmk,
\eea
where the slow-roll parameters are given by $\epsilon = (d \ln
f/d\Phi)^2,~\eta = d^2 \ln f/d\Phi^2+2\epsilon$. Here and hereafter, we
assume that the K\"ahler potential only depends on the combination
$XX^{\ast}$, though more general case makes $X$ less stable. The
effective mass squared along the inflationary trajectory is positive if
the following conditions are satisfied,
\beq
  K_{\Phi\Phi^{\ast}XX^{\ast}} + \frac{\eta}{2} - \epsilon \le 1,~~~~
  K_{XXX^{\ast}X^{\ast}} - \epsilon \le 0.
\eeq
More stringent conditions that the effective mass squared along the
inflationary trajectory is larger than the Hubble parameter squared,
which quickly drive Im $\Phi$ and $X$ to zeroes, require that
\beq
  K_{\Phi\Phi^{\ast}XX^{\ast}} \lesssim \frac56,~~~~
  K_{XXX^{\ast}X^{\ast}}  \lesssim - \frac13.
\eeq
Thus, as long as we have K\"ahler potential satisfying the above
conditions, we can construct an arbitrary potential from the function
$f$ in supergravity.

\subsubsection{Topological inflation}

\label{subsubsec:Ftopological}

As another type of large field inflation, we consider topological
inflation \cite{Linde:1994hy} in this subsection. Though topological
inflation has a similar form of the potential as that of new inflation,
it has interesting features. First of all, different from other low
scale inflations, it is free from the initial value problem, that is,
why the initial field is homogeneous over the horizon scale and is
fine-tuned to small region. Therefore, as long as the Universe is open
at the beginning, topological inflation can occur without
fine-tuning. Observationally, it predicts significant amount of the
tensor-to-scalar ratio, which will be confirmed or ruled out in near
future. Thus, topological inflation is still attractive.

Topological inflation is realized by the symmetry breaking potential
such as
\beq
  V(\phi_i) = \lambda \lmk \phi^2 - \la \phi \ra^2 \rmk^2,
\eeq
where $\phi^2 = \sum_{i=1}^{n} \phi_i^2$ and $\lambda$ is a coupling
constant. Domain walls, strings, and monopoles would be formed for
$n=1,2,3$ unless inflation happens. In order to support inflation, the
VEV of $\phi$, $\la \phi \ra$, must be larger than unity. This can be
understood from the following simple discussion. The typical radius of
topological defect $R \sim 1/(\sqrt{\lambda} \la \phi \ra)$ is
determined by equating the gradient energy density $(\la \phi
\ra/R)^{2}$ and the potential energy density $\lambda \la \phi \ra^{4}$.
For topological inflation to occur, the typical radius $R$ must be
larger than the hubble radius given by $H^{-1} \sim 1/(\sqrt{\lambda}
\la \phi \ra^{2})$, which leads to the condition $\la \phi \ra \gtrsim
1$. In fact, numerical calculations gives more precise conditions $\la
\phi \ra \gtrsim 1.7$ \cite{Sakai:1995nh} for a double-well potential
and $\la \phi \ra \gtrsim 0.95$ \cite{Kawasaki:2000tv} for the
supergravity model \cite{Izawa:1998rh}. Therefore, the exponential
factor in F-term potential is again problematic in constructing
topological inflation model in supergravity. Then, we give a model given
in Ref. \cite{Kawasaki:2001as}, in which the Nambu-Goldstone-like shift
symmetry is used to avoid the above problem.

Let us consider the following K\"ahler potential,
\beq
  K = - \frac12\,(\Phi - \Phi^{\ast})^{2} + |X|^{2},
  \label{eq:topoKahler}
\eeq
which is invariant under the shift symmetry,
\beq
  \Phi \longrightarrow \Phi + C
\eeq
with $C$ a real parameter. As small breaking of the shift
symmetry, we introduce the following superpotential,
\beq
  W = X (v - u \Phi^2) = v X (1 - g \Phi^2),
\eeq
which is invariant under the $R$ symmetry with $R$ charges of $X$ and
$\Phi$ to be $2$ and $0$, and under the $Z_2$ symmetry with $X$ and
$\Phi$ to be even and odd, respectively. While $v$ is of the order of
unity, $u \ll 1$ and $g \equiv u/v \ll 1$, representing the small
breaking of the shift symmetry. The Lagrangian density $\CL(\Phi,X)$ for
the scalar fields $\Phi$ and $X$ is given by
\beq
  \CL(\Phi,X) = - \partial_{\mu}\Phi\partial^{\mu}\Phi^{\ast} 
       - \partial_{\mu}X\partial^{\mu}X^{\ast}
         -V(\Phi,X),
  \label{eq:toplagrangian}
\eeq
with the scalar potential $V$ given by
\beq
  V = v^{2} e^{K} \lkk\,
      \left|\,1 - g\Phi^{2}\,\right|^{2}(1-|X|^{2}+|X|^{4}) 
       + |X|^{2} \left
         |\,2g\Phi + (\Phi-\Phi^{\ast})(1-g\Phi^{2})\,
                 \right|^{2}
                  \,\rkk.
    \label{eq:toppotential}
\eeq
By decomposing the scalar field $\Phi$ into the real component $\phi$
and the imaginary one $\chi$,
\beq
  \Phi = \frac{1}{\sqrt{2}}\,(\phi + i \chi),
  \label{eq:decomposition}
\eeq
we can easily show that the effective mass squared of $\chi$ becomes
$m_{\chi}^2 \simeq 6 H^2$, which quickly drives $\chi$ to zero. On the
other hand, though the effective mass of $X$ is not larger than the
Hubble parameter,\footnote{Considering higher order terms in the
K\"ahler potential can give $X$ an additional mass larger than the
Hubble parameter, which makes $X$ go to zero quickly.} it is clear that
$X$ is irrelevant for the dynamics and the primordial density
fluctuation, and hence $X$ can be safely set to zero. The potential for
the inflaton $\phi$ reduces to
\bea
  V &=& v^{2} \lmk 1 - \frac{g}{2} \phi^{2} \rmk^{2} \non \\ 
    &\simeq& v^{2} \lmk\,1 - g \phi^{2} \rmk  
    \quad~~ {\rm for} \quad \phi \lesssim 1. 
\eea
Then, the $e$-folding number $N$ is given by
\beq
  N \simeq \int_{\phi_{f}}^{\phi_{N}} \frac{V}{V'} 
    \simeq \frac{1}{ 2g } 
              \ln \lmk \frac{\phi_{f}}{\phi_{N}} \rmk, 
  \label{eq:topefold}
\eeq
which gives $\phi_{N} \sim \phi_{f} e^{-2 g N} \sim
\sqrt{\frac{2}{g}}\,e^{-2 g N}$.  Here $\phi_{f} \sim \sqrt{2/g}$ is the
value of $\phi$ at the end of topological inflation. On the other hand,
the slow-roll parameters are estimated as
\beq
  \epsilon \simeq 2 g^2 \phi_N^2 \simeq 4 g e^{-4 g N},~~~~
  \eta \simeq - 2 g.
\eeq
Then, the observable quantities are given by
\bea
  \CR^2 &=& \frac{1}{24\pi^2}\frac{V}{\epsilon} 
        = \frac{v^2}{48\pi^2 g^2 \phi_N^2}
        \simeq \frac{v^2}{96\pi^2 g} e^{4 g N}, \non \\ 
  n_s - 1 &=& -6\epsilon+2\eta \simeq 2\eta = -4 g, \non \\
  r &=& 16 \epsilon = 32 g^2 \phi_N^2 \simeq 64 g e^{-4 g N}.
  \label{eq:topcobs}
\eea
For $g = 0.01$ and $N = 60$, $n_s = 0.96$, $v \simeq 8.3 \times
10^{-5}$, and $\phi_N \simeq 4.3$. This model also predicts significant
amount of the tensor perturbation $r \simeq 0.06$ for $g = 0.01$, which
will be confirmed or ruled out in near future.

After topological inflation ends, the inflaton rapidly oscillates around
the global minimum $\la \phi \ra \equiv \pm \sqrt{2/g}$ and decays into
the standard particles to reheat the universe. The decay of the
inflaton can take place if we consider higher order terms $u'(\Phi^{2} +
\Phi^{\ast\,2}) |\psi_{i}|^{2}$ in the K\"ahler potential. Here $u'$ is
a constant associated with the breaking of the shift symmetry with
$\CO(u) = \CO(u')$ and $\psi_{i}$ are the standard particles. The
decay rate of the inflaton becomes $\Gamma \sim u'^{2} \la \phi \ra^2
m_\phi^3 \sim u'^2 \sqrt{g} v^3$ with $\la \phi \ra^2 = 2/g$ and
$m_{\phi} \simeq 2 \sqrt{g} v$. Then, the reheating temperature is given
by
\beq
  T_R \simeq \lmk \frac{90}{\pi^2 g_{\ast}} \rmk^{\frac14} \sqrt{\Gamma}
      \sim u' g^{\frac14} v^{\frac32}
      \sim g^{\frac54} v^{\frac25},
\eeq
where we have used $u' \sim u = g v$. For $g = 0.01$, $T_R \sim 2.0
\times 10^{-13} \sim 4.8 \times 10^{5}$ GeV.\footnote{Once the inflaton
acquires a non-vanishing VEV, it can decay into the SM particles through
supergravity effects \cite{Endo:2006qk}, which predicts higher reheating
temperature in this model.}

Finally, we comment on natural inflation as another type of large field
inflation, which was proposed in Ref. \cite{Freese:1990rb}. It is
attractive in that it can predict significant amount of tensor
fluctuations and the initial value problem is less severe. Such a model
in context of supergravity is discussed in Ref. \cite{German:2001sm},
for example.

\subsection{D-term inflation}

\label{subsec:D-term}

In the previous subsection, we have discussed inflation models based on
the F-term potential. The main obstacle to construct inflation model in
supergravity comes from F-term, specifically, the exponential factor
appearing in it. Therefore, if we can obtain positive potential energy
in D-term, it can lead to successful inflation without the $\eta$
problem, which was first pointed out by Stewart
\cite{Stewart:1994ts}. In this subsection, we give concrete examples of
inflation model supported by D-term.

\subsubsection{Hybrid inflation}

\label{subsubsec:Dhybrid}

Let us consider a D-term model of hybrid inflation proposed in
Ref. \cite{Binetruy:1996xj}. (See also Ref. \cite{Halyo:1996pp}.) We
introduce the following superpotential,
\beq
  W = \lambda S \Phi_{+} \Phi_{-},
\eeq
where $S$, $\Phi_{+}$, and $\Phi_{-}$ are three (chiral) superfields,
and $\lambda$ is a coupling constant. This superpotential is invariant
under a $U(1)$ gauge symmetry, whose charges are assigned to be $0, +1,
-1$ for the fields $S, \Phi_{+}, \Phi_{-}$,
respectively.\footnote{Strictly speaking, we need to modify the
assignment of the charges $q_{+}$ and $q_{-}$ for $\Phi_{+}$ and
$\Phi_{-}$ such that $q_{+} = 1-\xi/2$ and $q_{-} = -1-\xi/2$ for the
non-vanishing FI term $\xi$ in supergravity
\cite{Binetruy:2004hh}. However, since $\xi \ll 1$ as shown later,
$q_{+}$ and $q_{-}$ are approximated as $q_{+} \simeq 1$ and $q_{-}
\simeq -1$.} It also possesses the $R$ symmetry, under which they are
transformed as
\beq
  S(\theta) \longrightarrow e^{2i\alpha} S(\theta e^{i\alpha}), ~~~~~ 
  \Phi_{+}\Phi_{-}(\theta) \longrightarrow \Phi_{+}\Phi_{-}(\theta e^{i\alpha}).  
\eeq
We take the canonical K\"ahler potential invariant under the gauge and
the $R$ symmetries,
\beq
  K = |S|^2+|\Phi_{+}|^2+|\Phi_{-}|^2. 
  \label{eq:DHkahler}
\eeq
The tree level scalar potential is given by the standard formulae
(\ref{eq:Fpotential}) and (\ref{eq:Dpotential}),
\bea
  V(S,\Phi_{+},\Phi_{-}) &=& \lambda^2 e^{|S|^2+|\Phi_{+}|^2+|\Phi_{-}|^2}
          \lkk \left| \Phi_{+} \Phi_{-} \right|^2 
             + \left| S \Phi_{-} \right|^2 + \left| S \Phi_{+} \right|^2 
             + \lmk |S|^2 + |\Phi_{+}|^2+|\Phi_{-}|^2 + 3 \rmk 
                 \left| S \Phi_{+} \Phi_{-} \right|^2 \rkk \non \\
     && +~\frac{g^2}{2} \lmk |\Phi_{+}|^2-|\Phi_{-}|^2+\xi \rmk^2,
 \label{eq:DHpotential}
\eea
where $g$ is a gauge coupling constant, $\xi > 0$ is a non-vanishing FI
term, and we have taken a minimal gauge kinetic function. This potential
possesses the unique global minimum $V=0$ at
\beq
  S = \Phi_{+} = 0,~~~~~\Phi_{-} = \sqrt{\xi}. 
\eeq
However, for a large value of $|S|$, the potential has a local minimum
with positive energy density at
\beq
  \Phi_{+} = \Phi_{-} = 0.
\eeq
In order to find the critical value $S_c$ of $|S|$, we calculate the
mass matrix of $\Phi_{+}$ and $\Phi_{-}$ at the inflationary trajectory
$\Phi_{+} = \Phi_{-} = 0$, which is given by
\beq
  V_{\rm mass} = m_{+}^2 |\Phi_{+}|^2 + m_{-}^2 |\Phi_{-}|^2,
\eeq
with
\beq
  m_{+}^2 = \lambda^2 |S|^2 e^{|S|^2} + g^2 \xi,~~~~~~
  m_{-}^2 = \lambda^2 |S|^2 e^{|S|^2} - g^2 \xi.
\eeq
Thus, as long as $m_{-}^2 \ge 0$, which is equivalent to $|S| \ge S_c
\simeq g \sqrt{\xi} / \lambda$ for $S_c \lesssim 1$, the local minimum
$\Phi_{+} = \Phi_{-} = 0$ is stable so that inflation is driven by the
positive potential energy density $g^2 \xi^2 /2$. In addition, such a
mass split generates quantum correction calculated by the standard
formula \cite{Coleman:1973jx},
\bea
 V_{1L} &=& \frac{1}{32\pi^2}\lkk 
          (\lambda^2 |S|^2 e^{|S|^2} +g^2\xi)^2
          \ln \lmk \frac{\lambda^2 |S|^2 e^{|S|^2} +g^2\xi}{\Lambda^2} \rmk
        + (\lambda^2 |S|^2 e^{|S|^2} -g^2\xi)^2
          \ln \lmk \frac{\lambda^2 |S|^2 e^{|S|^2} -g^2\xi}{\Lambda^2} \rmk
          \right. \non \\ 
        && \quad\quad\quad  \left.
        - 2 \lambda^4 |S|^4 e^{2|S|^2} 
          \ln \lmk \frac{\lambda^2 |S|^2 e^{|S|^2}}{\Lambda^2} \rmk \rkk,
  \label{eq:Doneloop}
\eea
where $\Lambda$ is some renormalization scale. When $|S| \gg S_c$,
it is approximated as
\beq
 V_{1L} \simeq \frac{g^4 \xi^2}{16\pi^2} 
                \lkk \ln \lmk \frac{\lambda^2 |S|^2 e^{|S|^2}}{\Lambda^2} \rmk
                    + \frac32
                \rkk.
 \label{eq:DLapprox}
\eeq
Thus, the effective potential of $S$ during inflation is given by
\beq
 V(S) \simeq \frac{g^2\xi^2}{2} 
             \lkk 1 + \frac{g^2}{8\pi^2} 
              \ln \lmk \frac{\lambda^2 |S|^2 e^{|S|^2}}{\Lambda^2} \rmk
             \rkk.
\label{eq:DHpotea}
\eeq  
Since the above potential does not depend on the phase of the complex
scalar field $S$, we identify its real part $\sigma \equiv \sqrt{2} {\rm
Re} S$ with the inflaton without loss of generality. Then, for $\sigma_c
\ll \sigma \lesssim 1$, the effective potential of the inflaton $\sigma$
during inflation is given by
\beq
 V(\sigma) \simeq \frac{g^2\xi^2}{2} 
             \lkk 1 + \frac{g^2}{8\pi^2} 
              \ln \lmk \frac{\lambda^2 \sigma^2}{2\Lambda^2} \rmk
             \rkk.
\label{eq:DHpotea2}
\eeq  
The slow-roll parameters are estimated as
\beq
  \epsilon \simeq \frac{g^4}{32\pi^4\sigma^2},~~~~
  \eta \simeq - \frac{g^2}{4\pi^2\sigma^2}.
\eeq
The inflation ends if the inflaton $\sigma$ reaches $\sigma_c \equiv
\sqrt{S_c}$ or $\sigma_f \equiv g/(2\pi)$ corresponding to $|\eta|=1$.
The e-folding number is given by
\beq
  N \simeq \int_{\sigma_e}^{\sigma_N} \frac{4\pi^2 \sigma}{g^2} d\sigma
    \simeq \frac{2\pi^2}{g^2} \lmk \sigma_N^2 - \sigma_e^2 \rmk,
\eeq
where $\sigma_e = {\rm max}~(\sigma_c,\sigma_f)$. In the case that the
coupling $\lambda$ is not too small and $\lambda > 2\pi \sqrt{2\xi} \sim
0.01$, $\sigma_e = \sigma_f$. In this case, $\sigma_N^2 \simeq g^2 N
/(2\pi^2)$, $\epsilon \simeq g^2/(16\pi^2 N)$, and $\eta \simeq -1/(2
N)$. Then, the observable quantities are expressed as
\bea
  \CR^2 &=& \frac{1}{24\pi^2}\frac{V}{\epsilon} 
        \simeq \frac{N}{3} \xi^2, \non \\ 
  n_s - 1 &=& -6\epsilon+2\eta \simeq 2\eta 
          \simeq - \frac{1}{N}, \non \\
  r &=& 16 \epsilon \simeq \frac{g^2}{\pi^2 N}.
  \label{eq:Dhybridobs}
\eea
Inserting $\CR^2 \simeq 2.441 \times 10^{-9}$ and $N = 60$ yields $\xi
\simeq 1.1 \times 10^{-5}$, namely, $\sqrt{\xi} \simeq 3.3 \times 10^{-3}
\simeq 8.0 \times 10^{15}$~GeV. On the other hand, the spectral index
becomes $n_s \simeq 0.98$ for $N=60$ and hence is just outside the
observed values. However, cosmic strings are always formed at the end of
inflation because the $U(1)$ gauge symmetry is broken. Such cosmic
strings can contribute to the CMB anisotropies, but their contributions
cannot exceed $\sim 10$\% \cite{Bouchet:2000hd,Endo:2003fr}. Thus, the
model parameters are severely constrained. The detailed calculations
\cite{Rocher:2004et,Rocher:2004my} gives the following constraints,
\beq
  g \lesssim 2 \times 10^{-2} \quad {\rm and} \quad
  \lambda \lesssim 3 \times 10^{-5},
\eeq
which is equivalent to $\sqrt{\xi} \lesssim 2 \times
10^{15}$~GeV.\footnote{Recent results give a bit stronger constraint
\cite{Battye:2010xz}.} This constraint can be relaxed by introducing a
non-minimal K\"ahler potential \cite{Seto:2005qg} or curvaton mechanism
\cite{Endo:2003fr}, for example.

\subsubsection{Chaotic inflation}

\label{subsubsec:Dchaotic}

Though almost all of the inflation models based on D-term potential
belong to hybrid inflation type, it was recently pointed out that
chaotic inflation can be realized by D-term potential as well
\cite{Kadota:2007nc}. Here, following Ref. \cite{Kadota:2008pm}, we show
how to accommodate chaotic inflation in D-term potential.

Let us consider the following superpotential,
\beq
  W = \lambda S (X \overline{X} - \mu^2),
  \label{eq:Dchaosuper}
\eeq
where we have introduced three superfields $S, X, \overline{X}$ charged
under $U(1)$ gauge symmetry and (global) $U(1)_R$ symmetry. The $U(1)$
gauge charges of $S, X, \overline{X}$ are $0, +1, -1$ and their $R$
charges are $+2, 0, 0$, respectively. We set the constants $\lambda$ and
$\mu$ to be real and positive for simplicity. Taking the canonical
K\"ahler potential $K = |S|^2 + |X|^2 + |\overline{X}|^2$ and the
minimal gauge kinetic function $f_{a}(\Phi_i) = 1$, the
scalar potential reads
\bea
  V &=& V_F + V_D, \non \\ 
  V_F  &=&
       \lambda^2 e^{K} \lkk \,
         \biggl| X\overline{X} - \mu^2 \biggr|^2
         (1 - |S|^2 + |S|^4) \right. \non \\
       && \left.
          + |S|^2 \lhk
          \biggl| \overline{X} + X^{\ast}(X\overline{X} - \mu^2)  \biggr|^2
          + \left| X + \overline{X}^{\ast}(X\overline{X} - \mu^2)  \right|^2
               \rhk \,
            \rkk
        , \non \\
  V_D&=& \frac{g^2}{2} 
       \lmk |X|^2 - |\overline{X}|^2 \rmk^2,
  \label{eq:Dcahoscalarpot}
\eea
where $g$ is the coupling constant of the U(1) gauge symmetry. Note that
we do not need to introduce a non-vanishing FI term. The minima of the
F-term (the F-flat condition, $V_F=0$) are given by
\beq
        X\overline{X} - \mu^2 = 0 \quad {\rm and} \quad S = 0,
\eeq
and the minima of the D-term (the D-flat condition, $V_D=0$) are given
by
\beq
    |X| = |\overline{X}|.
\eeq
Then, the potential takes the global minima at
\beq
  S = 0,\,\,\, X = \mu e^{i\theta},\,\,\, \overline{X} = \mu e^{-i\theta},
\eeq
where the phase $\theta$ can be set to zero by the U(1) gauge
transformation.

Notice that this superpotential (\ref{eq:Dchaosuper}) and the
corresponding scalar potential (\ref{eq:Dcahoscalarpot}) are the same as
those (\ref{eq:Hybridsuper}) and (\ref{eq:Hpotential}) in the F-term
hybrid inflation model with the gauge symmetry $G$ to be Abelian, which
was discussed in the previous subsection. Here, the letters are changed
from $\Psi,\overline{\Psi}$ to $X,\overline{X}$ and $\mu^2$ is
rescaled. In the F-term hybrid inflation, the gauge singlet field $S \ll
1$ plays the role of an inflaton while $X$ and $\overline{X}$ remain
zeroes due to the heavy masses during the inflation, which satisfy the
D-flat condition. After inflation ends, they are destabilized and roll
down to the global minima. In order for this hybrid inflation to start,
the initial condition is such that the field $S$ has to be relatively
large but smaller than unity due to the exponential factor in F-term
while $X$ and $\overline{X}$ needs to almost vanish. On the other hand,
in this D-term chaotic inflation model, we consider another initial
condition given by $|X|\,\gtrsim\,1$ or $|\overline{X}|\,\gtrsim\,1$
with $S \sim 0$ and $X\overline{X} \sim \mu^2$, which almost satisfy the
F-flat condition. When the universe starts around the Planck scale, the
potential energy as well as the kinetic energy is expected to be of
order the Planck energy density.  Then, the almost F-flat direction is
naturally realized around the Planck scale due to the exponential factor
of F-term, which leads to the domination of the D-term potential so that
chaotic inflation takes place.

As explained above, the inflationary trajectory is expected to be given
by the (almost) F-flat direction, $S=0$ and $X \overline{X} = \mu^2$.
In fact, the effective mass squared of $S$ along this direction becomes
$\lambda^2 e^{K} (|X|^2 + |\overline{X}|^2)$ and hence is much larger
than the Hubble parameter squared $H^2 \simeq g^2 |X|^4/2$. Thus, we can
safely set $S$ to be zero and the potential reduces to
\beq
   V = \frac{\lambda^2}{4} e^{\frac12 (X^2 + \overline{X}^2)} 
        \lmk X\overline{X} - \mu'^2 \rmk^2 
      + \frac{g^2}{8} \lmk X^2 - \overline{X}^2 \rmk^2,
  \label{eq:Dchaoeffpot}
\eeq
where $\mu' \equiv \sqrt{2} \mu$ and we have redefined the fields $X
\equiv \sqrt{2}\, {\rm Re}\,X$, $\overline{X} \equiv \sqrt{2}\, {\rm
Re}\,\overline{X}$ (we take both $X$ and $\overline{X}$ to be positive
for definiteness). The kinetic terms of the fields $X$ and
$\overline{X}$ are still canonical. Though the trajectory is along the
valley of the two-dimensional configuration and is a bit complicated,
the F-flat condition $ X\overline{X} - \mu'^2 = 0$ is satisfied for $X
\gg 1$ because of the exponential factor. By inserting this condition
into Eq. (\ref{eq:Dchaoeffpot}), we obtain the the reduced potential
$V(X)$,
\beq
  V(X) \simeq \frac{g^2}{8} X^4.  
\eeq
As explicitly shown in Ref. \cite{Yamaguchi:2005qm}, when there is only
one massless mode and the other modes are massive, the generation of
adiabatic density fluctuations as wells as the dynamics of the
homogeneous mode is completely determined by this reduced potential
$V(X)$. Thus, this model of chaotic inflation has a quartic potential.

As given in Eq. (\ref{eq:chaoticobs}), the present constraint on the
tensor-to-scalar ratio $r < 0.24$ rules out the case with $n \ge 4$ if
the primordial density fluctuation is mainly generated by the inflaton.
However, if we take the non-minimal gauge kinetic function such as a
form $f = 1 + d_X |X|^2 + d_{\overline{X}} |\overline{X}|^2$
($d_X,d_{\overline{X}}$ : constants), chaotic inflation with a quadratic
potential is realized. Another D-term chaotic inflation with a quadratic
potential is also considered by use of the FI field
\cite{Kawano:2007gg}. Note that, in the model with the potential
(\ref{eq:Dcahoscalarpot}), no cosmic string is formed after inflation
because the U(1) gauge symmetry is already broken during inflation,
while the formation of cosmic strings severely constrain both of the
D-term and the F-term hybrid inflation models.

After the inflation, the inflaton starts to oscillate around the global
minimum and decays into standard particles. In this model, the inflaton
can decay into the right handed neutrino $N$, which quickly decays into
the standard particles through the Yukawa coupling. By introducing the
following superpotential,
\beq
  W = \alpha X \overline{X} N N,
 \label{eq:Nsuperpotential}
\eeq
the decay rate of the inflaton to the right handed neutrino is then given by
\bea
  \Gamma
     &\simeq& 
        \frac{1}{32\pi}
        \alpha^{2} \la X \ra^2 m
     \sim 
        \frac{1}{32\pi} \alpha^{2} \lambda \mu'^3 \non \\
     &\sim& 10^{-3}~{\rm GeV} 
          \lmk \frac{\alpha}{0.1} \rmk^{2}
          \lmk \frac{\lambda}{10^{-4}} \rmk
          \lmk \frac{\mu'}{10^{14}~\rm GeV} \rmk^3,
  \label{eq:Xdecay}
\eea
where $\alpha$ is the coupling constant of order unity, $\la X \ra
\simeq \mu'$ is the VEV of $X$, and $m \simeq \lambda \mu'$ is the mass
around the minimum. Then, the reheating temperature $T_R$ becomes
\beq
  T_R \simeq \lmk \frac{90}{\pi^2 g_{\ast}} \rmk^{\frac14} \sqrt{\Gamma}
      \sim 10^{7}~{\rm GeV} 
           \lmk \frac{\alpha}{0.1} \rmk
           \lmk \frac{\lambda}{10^{-4}} \rmk^{\frac12}
           \lmk \frac{\mu'}{10^{14}~\rm GeV} \rmk^{\frac32}.
\eeq
In this model, we can also show that the decay rate asymmetry of the
right handed neutrino to the standard particles generates lepton
asymmetry \cite{Fukugita:1986hr}, which is converted to baryon asymmetry
through sphaleron effects.

\section{Higgs inflation in Jordan frame supergravity}

\label{sec:Higgsinflation}

In this section, we discuss inflation models in Jordan frame
supergravity, focusing on Higgs chaotic inflation proposed a couple
years ago. The (classical) potential of the physical Higgs field $h$ is
given by $V(h) = (\lambda/4) (h^2 - v^2)^2 \sim \lambda h^4 /4$ for $h
\gg v$.  Then, this quartic type of potential can cause chaotic
inflation for $h \gg 1$. Unfortunately, as shown in the previous
section, the coupling $\lambda$ of order unity predicts too large
density fluctuation and also the large tensor-to-scalar ratio prohibits
a potential with quartic power. However, these constraints rely on three
important assumptions: (i) the Higgs field minimally couples to gravity,
(ii) the kinetic term of the Higgs field is canonical, (iii) The
primordial curvature perturbation is dominantly produced by the Higgs
field. If we could relax one of these three conditions, the Higgs field
may be responsible for inflation. Recently, interesting possibility
relaxing the assumption (i) was proposed by Bezrukov and Shaposhnikov,
in which a non-minimal coupling of the Higgs field to gravity is
considered \cite{Bezrukov:2007ep}.\footnote{See references for other
possibilities without the assumption (ii)
\cite{Germani:2010gm,Nakayama:2010sk,Kamada:2010qe} or (iii)
\cite{Langlois:2004nn}.} Motivated by this work, several attempts were
made to realize Higgs chaotic inflation in Jordan frame supergravity
\cite{Einhorn:2009bh,Lee:2010hj,Ferrara:2010yw,Nakayama:2010ga}.  In
this section, we first give the basic formulae of inflation where the
inflaton is non-minimally coupled to gravity, and explain how Higgs
inflation with quartic potential can circumvent the above constraints in
Jordan frame. Then, we give the formulation of a scalar field in Jordan
frame supergravity and show how to accommodate Higgs chaotic inflation
in this framework.

\subsection{Inflation in the Jordan frame}

\label{subsec:Jinflation}

In this subsection, following to Refs. \cite{Chiba:2008ia}, we derive
the slow-roll conditions for inflation in the Jordan frame and express
observational quantities in terms of the slow-roll parameters. The
action in the Jordan frame with metric $g_{\mu\nu}$ is given by
\beq
  S = \int d^4 x \sqrt{-g} \lmk \frac{\Omega(\phi)}{2}R  
       - \frac12 g^{\mu\nu} \del_{\mu} \phi \del_{\nu} \phi - V(\phi)
  \rmk,~~~~~
      \Omega(\phi) \equiv 1 - 2 F(\phi),
  \label{eq:Jaction}
\eeq
where $\phi$ is an inflaton, $\Omega(\phi)$ is a conformal factor, and
$F(\phi) R$ stands for a non-minimal coupling of the inflaton to
gravity. The case with $F = 0$ ($\Omega = 1$) corresponds to a minimal
coupling and the case with $F(\phi) = \phi^2/12$ corresponds to a
conformal coupling. Then, the equation of motion and the Friedmann
equation are given by
\bea
  && \Omega \lmk \ddot{\phi} + 3H \dot{\phi} + V' \rmk 
        + \frac{3 \Omega'}{2} \lmk \ddot{\Omega} + 3H \dot{\Omega} \rmk
        + \Omega' \lmk \frac12 \dot{\phi}^2 - 2V \rmk=0, \\
  && H^2 \Omega + H \dot{\Omega} = \frac13 \lkk \frac12 \dot{\phi}^2 + V(\phi) \rkk,
\eea
where we have assumed the flat Friedmann background and the
homogeneity of the scalar field. These equations are approximated as
\bea
  && 3 H \dot{\phi} \simeq 
      - \frac{\Omega^2}{f} \lmk \frac{V}{\Omega^2} \rmk' 
      \equiv - V_{\rm eff}',~~~~~~~
       f(\phi) \equiv
          1+\frac{3\Omega'^2(\phi)}{2 \Omega(\phi)}, \non \\ 
  && H^2 \Omega \simeq \frac{V}{3},
  \label{eq:Jsleq}
\eea
as long as the following three slow-roll conditions are
satisfied,\footnote{Strictly speaking, one subsidiary condition $|V_{\rm
eff}'/V'| = \CO(1)$ is also necessary.}
\bea
  && \epsilon_{J} \equiv \frac{\Omega V_{\rm eff}'^2}{2V^2};~~~~
       \epsilon_{J} \ll 1,
     \label{eq:Jslowroll1}\\
  && \eta_{J} \equiv \frac{\Omega V_{\rm eff}''}{V};~~~~|\eta_{J}|\ll 1,
     \label{eq:Jslow-roll2}\\
  && \delta_{J} \equiv \frac{\Omega' V_{\rm eff}'}{V};~~~~|\delta_{J}|\ll 1.
     \label{eq:Jslow-roll3}
\eea
Before going to observational quantities, we comment on the relation
between the Jordan frame and the Einstein frame. Introducing Einstein
metric $\widehat{g}_{\mu\nu}$ by conformal transformation with a
conformal factor $\Omega(\phi)$,
\beq
  \widehat{g}_{\mu\nu} = \Omega(\phi) g_{\mu\nu},
\eeq
the action (\ref{eq:Jaction}) in the Jordan frame can be rewritten in
the Einstein frame as
\beq
  S = \int d^4 x \sqrt{-\widehat{g}} \lmk \frac{\widehat{R}}{2} 
       - \frac12 \widehat{g}^{\mu\nu} 
          \del_{\mu} \widehat{\phi} \del_{\nu} \widehat{\phi} 
           - \widehat{V}(\widehat{\phi}) \rmk,
\eeq
where we have defined a scalar field $\widehat{\phi}$ with the canonical
kinetic term and its potential $\widehat{V}$,
\beq
  d\widehat{\phi}^2 \equiv \frac{f(\phi)}{\Omega(\phi)} d\phi^2,~~~~~
       \label{eq:corres1}
  \widehat{V}(\widehat{\phi})\equiv\frac{V(\phi)}{\Omega^2(\phi)}.
\label{eq:corres2}
\eeq
Notice that, in this section, the physical quantities in the Einstein
frame are characterized with hats while those in the Jordan frame
without hats. Then, it is easy to show that, as long as the slow-roll
conditions are satisfied, the slow-roll parameters in both frames are
related as
\beq
  \widehat{\epsilon} \equiv 
     \frac12 \lmk  \frac{1}{\widehat{V}} \frac{d\widehat{V}}{d\widehat{\phi}}
             \rmk^2
     \simeq \epsilon_{J} f,~~~~~
  \widehat{\eta} \equiv
      \frac{1}{\widehat{V}} \frac{d^2\widehat{V}}{d\widehat{\phi}^2}
     \simeq \eta_{J} - \frac32 \delta_{J} 
           + \frac12 \frac{f'}{f} \frac{\Omega}{\Omega'} \delta_{J},
\eeq
and that the curvature perturbation $\CR$ in the comoving gauge is
invariant under the conformal transformation, that is, $\widehat{\CR} =
\CR$. Then, the observable quantities are expressed as \cite{Makino:1991sg}
\bea
  \CR^2 &=& \frac{1}{24\pi^2}\frac{V}{\Omega^2\epsilon_{J} f} 
        ~~\lmk = \frac{1}{24\pi^2}\frac{\widehat{V}}{\widehat{\epsilon}} 
             = \widehat{\CR}^2 \rmk, \non \\ 
  n_s - 1 &=& - 6\epsilon_{J} f +2\eta_{J}-3\delta_{J} 
               + \frac{f'}{f} \frac{\Omega}{\Omega'} \delta_{J}
        ~~\lmk = - 6 \widehat{\epsilon} + 2 \widehat{\eta}
             = \widehat{n_s} - 1 ~\rmk, \non \\
  r &=& 16 \epsilon_{J} f 
        ~~\lmk = 16 \widehat{\epsilon} = \widehat{r} ~\rmk. 
  \label{eq:Jcobs}
\eea
The $e$-folding number $N$ is calculated as
\beq
  N = \int_{t_N}^{t_e} H dt \simeq \int_{\phi_e}^{\phi_N}
          \frac{V}{\Omega V'_{\rm eff}} d\phi
      ~~\lmk = \int_{\widehat{\phi}_e}^{\widehat{\phi}_N}
               \frac{\widehat{V}}{\frac{d\widehat{V}}{d\widehat{\phi}}} 
                d\widehat{\phi}
             = \widehat{N} ~\rmk.
\eeq

Now, we are ready for concrete examples. Let us consider chaotic
inflation with power-law potential $V(\phi) = \lambda_n \phi^n /n$, in
which the inflaton has a non-minimal coupling to gravity $F(\phi) = \xi
\phi^2/2$ ($\xi$ : a dimensionless parameter, $\Omega(\phi) = 1 - \xi
\phi^2$) \cite{Futamase:1987ua}. Then, the action is given by
\beq
  S = \int d^4x \sqrt{-g} \biggl[ 
       \frac12 \lmk 1 - \xi \phi^2 \rmk R 
       - \frac12 g^{\mu\nu} \del_{\mu} \phi \del_{\nu} \phi 
       - \frac{\lambda_n}{n} \phi^n
       \biggr]. 
  \label{eq:Jchaoaction}
\eeq
For $|\xi| \phi^2 \ll 1$, the slow-roll parameters and the $e$-folding
number are given by
\beq
   \epsilon_{J} = - \frac{n^2}{2\phi^2} 
       \simeq - \frac{n}{4N},~~~~
   \eta_{J} = \frac{n(n-1)}{\phi^2} \simeq \frac{n-1}{2N},~~~~
   \delta_{J} = -2n\xi,
\eeq
and
\beq
  N = \int_{\phi_e}^{\phi_N} \frac{V}{\Omega V'_{\rm eff}} d\phi
    \simeq \frac{\phi_N^2}{2n}.
\eeq
Thus, $|\xi| \ll 1$ and $\phi^2 \gg n$ are necessary for
slow-roll. Then, the observable quantities are evaluated as,
\bea
  \CR^2 &=& \frac{1}{24\pi^2} \frac{V}{\Omega^2 \epsilon_{J} f}
        \simeq \frac{\lambda_n}{12\pi^2} \frac{\phi_N^{n+2}}{n^3}, \non \\
  n_s - 1 &\simeq& -6\epsilon_{J} f+2\eta_{J}-3\delta_{J} 
        \simeq - \frac{n+2}{2N} + 6 n \xi, \non \\
  r &=& 16 \epsilon_{J} f \simeq \frac{4n}{N},
\eea
where we have used $f \simeq 1$. The tensor-to-scalar ratio $r$
coincides with that in the minimal coupling case. Thus, this case still
excludes chaotic inflation with $n \ge 4$ power-law index.

On the other hand, for $|\xi| \phi^2 \gg 1$ and $n \ne 4$, the slow-roll
parameters and the $e$-folding number are given by
\beq
   \epsilon_{J} = - \frac{(n-4)^2 \xi}{2(1-6\xi)^2},~~~~
   \eta_{J} = - \frac{(n-4)(n-1) \xi}{1-6\xi},~~~~
   \delta_{J} = - \frac{2(n-4)\xi}{1-6\xi},
\eeq
and
\beq
  N = \int_{\phi_e}^{\phi_N} \frac{V}{\Omega V'_{\rm eff}} d\phi
    \simeq \frac{1-6\xi}{|\xi|} \frac{1}{n-4} \ln \frac{\phi_N}{\phi_e}.
\eeq
Thus, $|\xi| \ll 1$ is required for slow-roll. Note that the positivity
of $\Omega$ coming from Eq. (\ref{eq:Jsleq}) requires $\xi < 0$. Then, the
observable quantities are evaluated as
\bea
  \CR^2 &=& \frac{1}{24\pi^2} \frac{V}{\Omega^2 \epsilon_{J} f}
        \simeq \frac{1}{12\pi^2} \frac{(1-6\xi)}{|\xi|^3} 
                \frac{\lambda_n}{n(n-4)^2} \phi^{n-4}, \non \\
  n_s - 1 &\simeq& -6\epsilon_{J} f +2\eta_{J}-3\delta_{J} 
        \simeq - \frac{|\xi|}{(1-6\xi)} (n-4)^2, \non \\
  r &=& 16 \epsilon_{J} f \simeq \frac{8 |\xi|}{(1-6\xi)} (n-4)^2
        \simeq 0.32 \lmk \frac{1-n_s}{0.04} \rmk,
\eea
where we have used $f \simeq 1- 6\xi \simeq 1$. Thus, the
tensor-to-scalar ratio may be too large in this case unless $1-n_s$
becomes smaller.

For $|\xi| \phi^2 \gg 1$ and $n = 4$, the slow-roll parameters and the
$e$-folding number are calculated as
\beq
   \epsilon_{J} = - \frac{8}{(1-6\xi)^2 \xi \phi^4} 
       \simeq - \frac{1}{8N^2\xi},~~~~
   \eta_{J} = \frac{4}{(1-6\xi) \phi^2} \simeq \frac{1}{2N},~~~~
   \delta_{J} = \frac{8}{(1-6\xi) \phi^2} \simeq \frac{1}{N},
\eeq
and
\beq
  N = \int_{\phi_e}^{\phi_N} \frac{V}{\Omega V'_{\rm eff}} d\phi
    \simeq \frac{1-6\xi}{8} \phi_N^2.
\eeq
Thus, the slow-roll conditions are always satisfied irrespective of
$\xi$, as long as $\xi<0$. This can be easily understood because
$V'_{\rm eff} \simeq 0$ and $d\widehat{V}/d\widehat{\phi} \simeq 0$ in
this case. Then, the observable quantities are expressed in terms of $N$
as,
\bea
  \CR^2 &=& \frac{1}{24\pi^2} \frac{V}{\Omega^2 \epsilon_{J} f}
        \simeq \frac{\lambda_4}{12\pi^2} \frac{N^2}{|\xi|(1-6\xi)}, \non \\
  n_s - 1 &\simeq& -6\epsilon_{J} f+2\eta_{J}-3\delta_{J} 
        \simeq - \frac{3(1-6\xi)}{4 N^2 |\xi|} - \frac{2}{N}, \non \\
  r &=& 16 \epsilon_{J} f \simeq \frac{2(1-6\xi)}{N^2 |\xi|},
\eea
where we have used $f \simeq 1- 6\xi$. Inserting $\CR^2 \simeq 2.441
\times 10^{-9}$ and $N=60$ yields the relation $\lambda_4 \simeq 4.8
\times 10^{-10} \xi^2$, namely, $|\xi| \simeq 4.5 \times 10^{4}
\sqrt{\lambda_4}$.\footnote{In fact, we need to take into account loop
effects, which are sensitive to the details of the UV completion
\cite{DeSimone:2008ei}.} In addition, $n_s \simeq 0.97$ and the
tensor-to-scalar ratio becomes $r \simeq 3.3 \times 10^{-3}$ in this
case. Thus, chaotic inflation with a quartic potential ($n=4$) can be
still viable for $|\xi| \phi^2 \gg 1$ in the Jordan frame, which
strongly motivates us to consider Higgs inflation non-minimally coupled
to gravity.

\subsection{Higgs Inflation in the Jordan frame supergravity}

\label{subsec:HJinflation}

In this subsection, we will first give the scalar part of the Lagrangian
in the Jordan frame supergravity and later discuss Higgs chaotic
inflation in it.

Since detailed derivation based on the superconformal supergravity by
gauge-fixing is given in Refs. \cite{Ferrara:2010yw}, we will only give
the relevant results. The scalar part of the Lagrangian in Jordan frame
supergravity is determined by the four functions, frame function
$\Omega(\Phi_i,\Phi^{\ast}_i)$, K\"ahler potential
$K(\Phi_i,\Phi^{\ast}_i)$ (independent of the frame function),
superpotential $W(\Phi_i)$, and gauge kinetic function $f(\Phi_i)$
\cite{Ferrara:2010yw}.  While the superpotential and the gauge kinetic
function are holomorphic functions of complex scalar fields, the frame
function and the K\"ahler potential are not holomorphic and real
functions of the scalar fields $\Phi_i$ and their conjugates
$\Phi_i^{\ast}$. The frame function $\Omega$ corresponds to a conformal
factor and could stand for scalar-gravity coupling $\Omega R /2$. In
particular, $\Omega = 1$ corresponds to the minimal coupling. Then, the
action of the scalar and the gravity sectors in the Jordan frame is
given by
\beq
  S = \int d^4x \sqrt{-g} \biggl[ 
       \frac{\Omega}{2} R + \frac{1}{\sqrt{-g}} \CL_{\rm kin} 
       - V(\Phi_i,\Phi_i^{\ast}) \biggr],
\eeq
where the kinetic terms of the scalar fields are determined by frame
function $\Omega$ and K\"ahler potential $K$ as
\beq
  \frac{1}{\sqrt{-g}} \CL_{\rm kin} =
      \lmk - \Omega K_{ij^{\ast}} 
        + 3\frac{\Omega_i \Omega_{j^{\ast}}}{\Omega} \rmk 
           \del_{\mu} \Phi_i \del_{\nu} \Phi_j^{\ast} g^{\mu\nu}
    - 3 \Omega \CA_{\mu}^2.
  \label{eq:Jkinetic}
\eeq
Here the lower indices of the frame function and the K\"ahler potential
represent the derivatives, and $\CA_{\mu}$ is the on-shell auxiliary
axial-vector field given by
\beq
  \CA_{\mu} = - \frac{i}{2\Omega} 
              \lmk \Omega_i \widetilde{\del_{\mu}} \Phi_i 
                 - \Omega_{i^{\ast}} \widetilde{\del}_{\mu} \Phi_i^{\ast}
              \rmk, 
\eeq
with $\widetilde{\del}_{\mu}$ to be the gauge covariant derivative. On
the other hand, the potential $V$ in the Jordan frame is related to the
potential $\widehat{V}$ in the Einstein frame and is given by
\beq
  V = \Omega^2 \widehat{V},~~~~~~~
   \widehat{V} = e^{K} \left[ D_{\Phi_i}W K_{ij^{\ast}}^{-1} D_{\Phi_j^{*}}W^{*} 
                     - 3 |W|^{2} \right] 
                  + \frac12 \sum_{a} \lkk {\rm Re} f_{a}(\Phi_i) \rkk^{-1} 
                     g_a^2 D_a^2.
\eeq  
Although frame function and K\"ahler function are independent in
general, we have a special class of the superconformal models, where the
following relation is satisfied
\beq
  \Omega(\Phi_i,\Phi_i^{\ast}) = e^{-\frac13 K(\Phi_i,\Phi_i^{\ast})}~~~~~ 
    \Longleftrightarrow~~~~~ 
  K(\Phi_i,\Phi_i^{\ast}) = - 3 \ln \Omega(\Phi_i,\Phi_i^{\ast}).
\eeq 
Then, the kinetic terms of the scalar fields in this case reduce to
\beq
  \frac{1}{\sqrt{-g}} \CL_{\rm kin} =
         3 \Omega_{ij^{\ast}}
           \del_{\mu} \Phi_i \del_{\nu} \Phi_j^{\ast} g^{\mu\nu}
         - 3 \Omega \CA_{\mu}^2.
  \label{eq:rJkinetic}
\eeq
You can easily find that the following form of the frame function,
assuming $\CA_{\mu}=0$, leads to the canonical kinetic terms of scalar
fields,
\beq
  \Omega(\Phi_i,\Phi_i^{\ast}) = 1 - 
    \frac13 \biggl[
              \delta_{ij} \Phi_i \Phi_j^{\ast} + J(\Phi_i) + J^{\ast}(\Phi_i^{\ast})
            \biggr] ,
\eeq
where $J(\Phi_i)$ is an arbitrary function. By taking this form of the
frame function and setting $\CA_{\mu} = 0$, the action of the scalar and
the gravity sectors in the Jordan frame reads
\beq
  S = \int d^4x \sqrt{-g} \biggl[ 
       \frac{\Omega}{2} R 
        - \delta_{ij} \del_{\mu} \Phi_i \del_{\nu} \Phi_j^{\ast} g^{\mu\nu} 
       - V(\Phi_i,\Phi_i^{\ast}) \biggr].
\eeq

Now, we give a model of Higgs inflation in the context of the Jordan
frame supergravity. As explicitly shown in Ref. \cite{Einhorn:2009bh},
as long as we take a superpotential with a form of $W = C + \mu H_u H_d$
($C, \mu$ : constants), the inflationary trajectory is unstable and/or
too steep. Then, we need to extend this superpotential by introducing
another superfield $S$, as done in the (F-term) chaotic inflation models
in the Einstein supergravity. Such an extention is accommodated in the
next-to-minimal supersymmetric standard model (NMSSM). Therefore,
chaotic inflation with NMSSM in the Jordan frame supergravity was
proposed in Ref. \cite{Einhorn:2009bh} and modified in
Refs. \cite{Lee:2010hj,Ferrara:2010yw}.\footnote{More general models of
inflation in the Jordan frame supergravity are discussed in
Refs. \cite{Nakayama:2010ga,Kallosh:2010ug}. In addition, another
interesting class of inflation in context of $F(R)$ supergravity is
considered in Refs. \cite{Kaneda:2010ut}.} We take the following frame
function, K\"ahler potential, and the superpotential,
\bea
  && \Omega = 1 
      - \frac13 \lmk |S|^2 + H_u H_u^{\dagger} + H_d H_d^{\dagger} \rmk
      - \frac12 \gamma (H_u H_d + {\rm h.c.}) + \frac{\zeta}{3} |S|^4,
  \non \\
  && K = -3 \ln \Omega, \non \\
  && W = - \lambda S H_u H_d + \frac{\rho}{3} S^3.
\eea
In fact, we can safely set the charged fields $H_u^{+}, H_d^{-}$ to be
zeroes, that is,
\beq
  H_u = \left( \begin{array}{c} 0 \\ H_u^0 \end{array} \right),~~~~~~
  H_d = \left( \begin{array}{c} H_d^0 \\ 0 \end{array} \right).
\eeq
With these truncations, three functions reduce to
\bea
  && \Omega = 1 
      - \frac13 \lmk |S|^2 + \left| H_u^0 \right|^2 + \left| H_d^0 \right|^2
                \rmk 
      + \frac12 \gamma (H_u^0 H_d^0 + H_u^{0 \ast} H_d^{0 \ast}) 
                         + \frac{\zeta}{3} |S|^4,
  \non \\
  && K = -3 \ln \Omega, \non \\
  && W = - \lambda S H_u H_d + \frac{\rho}{3} S^3,
\eea
which yield the D-term potential in the Jordan frame,
\beq
  V_D = \frac18 (g^2+g'^2) \lmk 
           \left| H_u^0 \right|^2 - \left| H_d^0 \right|^2
                           \rmk^2.
\eeq
Here we have taken a minimal gauge kinetic function $f_{a} = 1$, and $g$
and $g'$ are the gauge couplings of the $SU(2)_{\rm L}$ and the
$U(1)_{\rm Y}$ symmetries, respectively. Decomposing the complex scalar
fields into
\beq
  S = \frac{1}{\sqrt{2}} s e^{i\alpha},~~~~~
  H_u^0 = \frac{1}{\sqrt{2}} h \cos{\beta}\,e^{i\alpha_1},~~~~~ 
  H_d^0 = \frac{1}{\sqrt{2}} h \sin{\beta}\,e^{i\alpha_2}, 
\eeq
the D-term potential vanishes when $\beta = \pi/4$. Then, we take an
inflationary trajectory as $\beta=\pi/4$, $\alpha_i=0$, and $s=0$, whose
stability condition will be given below. Under these settings, the
action reduces to
\beq
  S = \int d^4x \sqrt{-g} \biggl[ 
       \frac12 \lmk 1 - \lhk \frac16-\frac{\gamma}{4} \rhk h^2 \rmk R 
       - \frac12 g^{\mu\nu} \del_{\mu} h \del_{\nu} h - \frac{\lambda^2}{16} h^4
       \biggr]. 
\eeq
This action is equivalent to the action (\ref{eq:Jchaoaction}) with
$n=4$, $\xi = 1/6- \gamma/4$, and $\lambda_4 = \lambda^2/4$. Thus,
chaotic inflation can be realized in the Higgs sector with NMSSM by use
of the Jordan frame supergravity. The detailed analysis, in fact, shows
that this inflationary trajectory is stable as long as $\zeta >
2|\lambda \rho|/\lambda^2 h^2 + 0.0327$ \cite{Ferrara:2010yw}.

After the inflation ends at $h \simeq (4/3)^{\frac14}/\sqrt{|\xi|}$, the
reheating quickly occurs because the interactions of the Higgs boson
with the other SM particles are strong. Then, the reheating temperature
$T_R$ is estimated as \cite{Bezrukov:2007ep}
\beq
  T_R \sim \left( \frac{10}{\pi^2 g_{\ast}} \right)^{\frac14} 
   \frac{\lambda_4^{\frac14}}{\sqrt{|\xi|}}
      \sim 3 \times 10^{15}\, {\rm GeV},
\eeq
where $g_{\ast} \simeq 200$ is the number of relativistic degrees of
freedom and we have used the relation $|\xi| \simeq 4.5 \times 10^{4}
\sqrt{\lambda_4}$.

\section{Conclusion}

\label{sec:conclusion}

In this paper, we have discussed inflation models in supergravity. After
explaining why it is difficult to accommodate inflation in supergravity,
we gave the prescriptions to circumvent such difficulties. Focusing on
the cases with almost canonical K\"ahler potential, we gave concrete
examples of each type of inflation. Though it was long supposed that it
was almost impossible to construct natural model of chaotic inflation,
we now have all types of inflation in supergravity. The ongoing
observations would confirm or exclude specific type of inflation. In
particular, chaotic inflation generates significant amount of primordial
tensor perturbations, which may be detected in near future. Then, next
step to develop inflation models in supergravity is to embed them in a
realistic model of particle physics like GUT and/or in a superstring
theory. It is interesting whether K\"ahler potential and superpotential
suitable for inflation naturally appears in the context of GUT and/or
superstring.

We have also discussed inflation models based on Jordan frame
supergravity, focusing on Higgs chaotic inflation. Since inflation
models in Jordan frame supergravity appeared very recently, we need to
investigate them in more detail, particularly paying attention to the
difference between inflation models in the Jordan frame and those in the
Einstein frame. We also should extend these models in the context of
superstring because a non-minimal coupling is naturally found in it.

Although $F(R)$ inflation models were also formulated in context of
$F(R)$ supergravity recently \cite{Gates:2009hu}, other important
classes of inflation such as $k$ inflation
\cite{ArmendarizPicon:1999rj}, ghost inflation
\cite{ArkaniHamed:2003uz}, DBI inflation \cite{Alishahiha:2004eh}, and
$G$ inflation \cite{Kobayashi:2010cm} are not yet formulated in
supergravity.\footnote{See Ref. \cite{Khoury:2010gb} for recent attempt
to supersymmetrize these higher derivative models of inflation.}
Supersymmetrization of these inflation models is also an important
topic.

\section*{Acknowledgments}

We would like to thank Takeshi Chiba, Kazuhide Ichikawa, Kohei Kamada,
Masahiro Kawasaki, Fuminobu Takahashi and Jun'ichi Yokoyama for useful
comments. This work is supported in part by JSPS Grant-in-Aid for
Scientific Research No .21740187.


\begin{thebibliography}{9}

\bibitem{WMAP}
  E.~Komatsu {\it et al.},
  arXiv:1001.4538 [astro-ph.CO].

\bibitem{inflation}
  A.~H.~Guth,
  Phys.\ Rev.\  {\bf D23}, 347-356 (1981);
  K.~Sato,
  Mon.\ Not.\ Roy.\ Astron.\ Soc.\  {\bf 195}, 467-479 (1981);
  A.~A.~Starobinsky,
  Phys.\ Lett.\  B {\bf 91}, 99 (1980).

\bibitem{Hmass}
  M.~Dine, W.~Fischler, D.~Nemeschansky,
  Phys.\ Lett.\  {\bf B136}, 169 (1984);
  G.~D.~Coughlan, R.~Holman, P.~Ramond, G.~G.~Ross,
  Phys.\ Lett.\  {\bf B140}, 44 (1984).

\bibitem{Copeland:1994vg}
  E.~J.~Copeland, A.~R.~Liddle, D.~H.~Lyth, E.~D.~Stewart and D.~Wands,
  Phys.\ Rev.\  D {\bf 49}, 6410 (1994)
  [arXiv:astro-ph/9401011].

\bibitem{Stewart:1994ts}
  E.~D.~Stewart,
  Phys.\ Rev.\  D {\bf 51}, 6847 (1995)
  [arXiv:hep-ph/9405389].


\bibitem{Murayama:1992ua}
  H.~Murayama, H.~Suzuki, T.~Yanagida and J.~Yokoyama,
  Phys.\ Rev.\ Lett.\  {\bf 70}, 1912 (1993);
  S.~Kasuya, T.~Moroi and F.~Takahashi,
  Phys.\ Lett.\  B {\bf 593}, 33 (2004)
  [arXiv:hep-ph/0312094];
  R.~Allahverdi, K.~Enqvist, J.~Garcia-Bellido and A.~Mazumdar,
  Phys.\ Rev.\ Lett.\  {\bf 97}, 191304 (2006)
  [arXiv:hep-ph/0605035];
  R.~Allahverdi, K.~Enqvist, J.~Garcia-Bellido, A.~Jokinen and A.~Mazumdar,
  JCAP {\bf 0706}, 019 (2007)
  [arXiv:hep-ph/0610134].


\bibitem{Bezrukov:2007ep}
  F.~L.~Bezrukov and M.~Shaposhnikov,
  Phys.\ Lett.\  B {\bf 659}, 703 (2008)
  [arXiv:0710.3755 [hep-th]];

\bibitem{Germani:2010gm}
  C.~Germani and A.~Kehagias,
  Phys.\ Rev.\ Lett.\  {\bf 105}, 011302 (2010)
  [arXiv:1003.2635 [hep-ph]];
  C.~Germani and A.~Kehagias,
  JCAP {\bf 1005}, 019 (2010)
  [Erratum-ibid.\  {\bf 1006}, E01 (2010)]
  [arXiv:1003.4285 [astro-ph.CO]].

\bibitem{Nakayama:2010sk}
  K.~Nakayama and F.~Takahashi,
  arXiv:1008.4457 [hep-ph].

\bibitem{Kamada:2010qe}
  K.~Kamada, T.~Kobayashi, M.~Yamaguchi and J.~Yokoyama,
  arXiv:1012.4238 [astro-ph.CO].

\bibitem{stringreview}
  L.~McAllister and E.~Silverstein,
  Gen.\ Rel.\ Grav.\  {\bf 40}, 565 (2008)
  [arXiv:0710.2951 [hep-th]];
2
  D.~Baumann and L.~McAllister,
  Ann.\ Rev.\ Nucl.\ Part.\ Sci.\  {\bf 59}, 67 (2009)
  [arXiv:0901.0265 [hep-th]].

\bibitem{SUGRAinflation}
  K.~A.~Olive,
  Phys.\ Rept.\  {\bf 190}, 307 (1990);
  D.~H.~Lyth and A.~Riotto,
  Phys.\ Rept.\  {\bf 314}, 1 (1999)
  [arXiv:hep-ph/9807278];
  D.~H.~Lyth,
  Lect.\ Notes Phys.\  {\bf 738}, 81 (2008)
  [arXiv:hep-th/0702128];
  A.~Mazumdar and J.~Rocher,
  arXiv:1001.0993 [hep-ph].


\bibitem{Bardeen:1980kt}
  J.~M.~Bardeen,
  Phys.\ Rev.\  D {\bf 22}, 1882 (1980).

\bibitem{pert} 
  S.~W.~Hawking,
  Phys.\ Lett.\  B {\bf 115}, 295 (1982);
  A.~A.~Starobinsky,
  Phys.\ Lett.\  B {\bf 117}, 175 (1982);
  A.~H.~Guth and S.~Y.~Pi,
  Phys.\ Rev.\ Lett.\  {\bf 49}, 1110 (1982).

\bibitem{Starobinsky:1979ty}
  A.~A.~Starobinsky,
  JETP Lett.\  {\bf 30}, 682 (1979)
  [Pisma Zh.\ Eksp.\ Teor.\ Fiz.\  {\bf 30}, 719 (1979)].

\bibitem{Lyth:2009zz}
  D.~H.~Lyth and A.~R.~Liddle,
  ``The primordial density perturbation: Cosmology, inflation and the origin of
  structure,''
{\it  Cambridge, UK: Cambridge Univ. Pr. (2009) 497 p}

\bibitem{SUSYreview}
For reviews of SUSY and supergravity, see
  J.~Wess and J.~Bagger,
  ``Supersymmetry and supergravity,''
{\it  Princeton, USA: Univ. Pr. (1992) 259 p};
  H.~P.~Nilles,
  Phys.\ Rept.\  {\bf 110}, 1 (1984);
  D.~Bailin and A.~Love,
  ``Supersymmetric gauge field theory and string theory,''
{\it  Bristol, UK: IOP (1994) 322 p. (Graduate student series in physics)}.

\bibitem{Berera:1995ie}
  A.~Berera,
  Phys.\ Rev.\ Lett.\  {\bf 75}, 3218 (1995)
  [arXiv:astro-ph/9509049].

\bibitem{BasteroGil:2006vr}
  M.~Bastero-Gil and A.~Berera,
  Phys.\ Rev.\  D {\bf 76}, 043515 (2007)
  [arXiv:hep-ph/0610343];
  M.~Bastero-Gil and A.~Berera,
  Int.\ J.\ Mod.\ Phys.\  A {\bf 24}, 2207 (2009)
  [arXiv:0902.0521 [hep-ph]].

\bibitem{Gaillard:1995az}
  M.~K.~Gaillard, H.~Murayama and K.~A.~Olive,
  Phys.\ Lett.\  B {\bf 355}, 71 (1995)
  [arXiv:hep-ph/9504307];
  S.~Antusch, M.~Bastero-Gil, K.~Dutta, S.~F.~King and P.~M.~Kostka,
  Phys.\ Lett.\  B {\bf 679}, 428 (2009)
  [arXiv:0905.0905 [hep-th]].

\bibitem{Stewart:1996ey}
  E.~D.~Stewart,
  Phys.\ Lett.\  B {\bf 391}, 34 (1997)
  [arXiv:hep-ph/9606241];
  E.~D.~Stewart,
  Phys.\ Rev.\  D {\bf 56}, 2019 (1997)
  [arXiv:hep-ph/9703232].

\bibitem{Kumekawa:1994gx}
  K.~Kumekawa, T.~Moroi and T.~Yanagida,
  Prog.\ Theor.\ Phys.\  {\bf 92}, 437 (1994)
  [arXiv:hep-ph/9405337].

\bibitem{Kawasaki:2006gs}
  M.~Kawasaki, F.~Takahashi and T.~T.~Yanagida,
  Phys.\ Lett.\  B {\bf 638}, 8 (2006)
  [arXiv:hep-ph/0603265].
  T.~Asaka, S.~Nakamura and M.~Yamaguchi,
  Phys.\ Rev.\  D {\bf 74}, 023520 (2006)
  [arXiv:hep-ph/0604132];
  M.~Endo, K.~Kadota, K.~A.~Olive, F.~Takahashi and T.~T.~Yanagida,
  JCAP {\bf 0702}, 018 (2007)
  [arXiv:hep-ph/0612263];
  M.~Endo, F.~Takahashi and T.~T.~Yanagida,
  Phys.\ Lett.\  B {\bf 658}, 236 (2008)
  [arXiv:hep-ph/0701042].

\bibitem{Endo:2006qk}
  M.~Endo, M.~Kawasaki, F.~Takahashi and T.~T.~Yanagida,
  Phys.\ Lett.\  B {\bf 642}, 518 (2006)
  [arXiv:hep-ph/0607170];
  M.~Endo, F.~Takahashi and T.~T.~Yanagida,
  Phys.\ Rev.\  D {\bf 76}, 083509 (2007)
  [arXiv:0706.0986 [hep-ph]].


\bibitem{Kawasaki:2004yh}
  For recent constraints, see
  M.~Kawasaki, K.~Kohri, and T.~Moroi,
  Phys.\ Lett.\  B {\bf 625}, 7 (2005)
  [arXiv:astro-ph/0402490];
  M.~Kawasaki, K.~Kohri, and T.~Moroi,
  Phys.\ Rev.\  D {\bf 71}, 083502 (2005)
  [arXiv:astro-ph/0408426];
  M.~Kawasaki, K.~Kohri, T.~Moroi and A.~Yotsuyanagi,
  Phys.\ Rev.\  D {\bf 78}, 065011 (2008)
  [arXiv:0804.3745 [hep-ph]].



\bibitem{new}
  A.~D.~Linde,
  Phys.\ Lett.\  B {\bf 108}, 389 (1982);
  A.~Albrecht and P.~J.~Steinhardt,
  Phys.\ Rev.\ Lett.\  {\bf 48}, 1220 (1982).

\bibitem{Izawa:1996dv}
  K.~I.~Izawa and T.~Yanagida,
  Phys.\ Lett.\  B {\bf 393}, 331 (1997)
  [arXiv:hep-ph/9608359].

\bibitem{Izawa:1997df}
  K.~I.~Izawa, M.~Kawasaki and T.~Yanagida,
  Phys.\ Lett.\  B {\bf 411}, 249 (1997)
  [arXiv:hep-ph/9707201].

\bibitem{chaonew}
  M.~Yamaguchi and J.~Yokoyama,
  Phys.\ Rev.\  D {\bf 63}, 043506 (2001)
  [arXiv:hep-ph/0007021];
  M.~Yamaguchi,
  Phys.\ Rev.\  D {\bf 64}, 063502 (2001)
  [arXiv:hep-ph/0103045];
  M.~Yamaguchi,
  Phys.\ Rev.\  D {\bf 64}, 063503 (2001)
  [arXiv:hep-ph/0105001].

\bibitem{double}
  M.~Kawasaki, N.~Sugiyama and T.~Yanagida,
  Phys.\ Rev.\  D {\bf 57}, 6050 (1998)
  [arXiv:hep-ph/9710259];
  M.~Kawasaki and T.~Yanagida,
  Phys.\ Rev.\  D {\bf 59}, 043512 (1999)
  [arXiv:hep-ph/9807544];
  T.~Kanazawa, M.~Kawasaki, N.~Sugiyama and T.~Yanagida,
  Phys.\ Rev.\  D {\bf 61}, 023517 (2000)
  [arXiv:hep-ph/9908350];
  T.~Kanazawa, M.~Kawasaki and T.~Yanagida,
  Phys.\ Lett.\  B {\bf 482}, 174 (2000)
  [arXiv:hep-ph/0002236];
  M.~Kawasaki, T.~Takayama, M.~Yamaguchi and J.~Yokoyama,
  Phys.\ Rev.\  D {\bf 74}, 043525 (2006)
  [arXiv:hep-ph/0605271];
  T.~Kawaguchi, M.~Kawasaki, T.~Takayama, M.~Yamaguchi and J.~Yokoyama,
  Mon.\ Not.\ Roy.\ Astron.\ Soc.\  {\bf 388}, 1426 (2008)
  [arXiv:0711.3886 [astro-ph]].


\bibitem{Asaka:1999yc}
  T.~Asaka, M.~Kawasaki and M.~Yamaguchi,
  Phys.\ Rev.\  D {\bf 61}, 027303 (2000)
  [arXiv:hep-ph/9906365].


\bibitem{hybrid}
  A.~D.~Linde,
  Phys.\ Lett.\  B {\bf 259}, 38 (1991);
  A.~D.~Linde,
  Phys.\ Rev.\  D {\bf 49}, 748 (1994)
  [arXiv:astro-ph/9307002].

\bibitem{GUTinf}
  G.~R.~Dvali, Q.~Shafi and R.~K.~Schaefer,
  Phys.\ Rev.\ Lett.\  {\bf 73}, 1886 (1994)
  [arXiv:hep-ph/9406319];

\bibitem{Stewart:1994pt}
  E.~D.~Stewart,
  Phys.\ Lett.\  B {\bf 345}, 414 (1995)
  [arXiv:astro-ph/9407040].

\bibitem{Lazarides:1995vr}
  G.~Lazarides and C.~Panagiotakopoulos,
  Phys.\ Rev.\  D {\bf 52}, R559 (1995)
  [arXiv:hep-ph/9506325].

\bibitem{Jeannerot:2000sv}
  R.~Jeannerot, S.~Khalil, G.~Lazarides and Q.~Shafi,
  JHEP {\bf 0010}, 012 (2000)
  [arXiv:hep-ph/0002151].

\bibitem{Antusch:2008pn}
  S.~Antusch, M.~Bastero-Gil, K.~Dutta, S.~F.~King and P.~M.~Kostka,
  JCAP {\bf 0901}, 040 (2009)
  [arXiv:0808.2425 [hep-ph]];
  S.~Antusch, K.~Dutta and P.~M.~Kostka,
  Phys.\ Lett.\  B {\bf 677}, 221 (2009)
  [arXiv:0902.2934 [hep-ph]];
  S.~Antusch, K.~Dutta and P.~M.~Kostka,
  AIP Conf.\ Proc.\  {\bf 1200}, 1007 (2010)
  [arXiv:0908.1694 [hep-ph]].

\bibitem{Linde:1997sj}
  A.~D.~Linde and A.~Riotto,
  Phys.\ Rev.\  D {\bf 56}, 1841 (1997)
  [arXiv:hep-ph/9703209].

\bibitem{Panagiotakopoulos:1997qd}
  C.~Panagiotakopoulos,
  Phys.\ Rev.\  D {\bf 55}, 7335 (1997)
  [arXiv:hep-ph/9702433];
  C.~Panagiotakopoulos,
  Phys.\ Lett.\  B {\bf 402}, 257 (1997)
  [arXiv:hep-ph/9703443].


\bibitem{Coleman:1973jx}
  S.~R.~Coleman and E.~J.~Weinberg,
  Phys.\ Rev.\  D {\bf 7}, 1888 (1973).


\bibitem{Jeannerot:2005mc}
  R.~Jeannerot and M.~Postma,
  JHEP {\bf 0505}, 071 (2005)
  [arXiv:hep-ph/0503146];
  M.~Bastero-Gil, S.~F.~King and Q.~Shafi,
  Phys.\ Lett.\  B {\bf 651}, 345 (2007)
  [arXiv:hep-ph/0604198];
  M.~ur Rehman, V.~N.~Senoguz and Q.~Shafi,
  Phys.\ Rev.\  D {\bf 75}, 043522 (2007)
  [arXiv:hep-ph/0612023];
  G.~Lazarides and C.~Pallis,
  Phys.\ Lett.\  B {\bf 651}, 216 (2007)
  [arXiv:hep-ph/0702260];
  M.~U.~Rehman, Q.~Shafi and J.~R.~Wickman,
  Phys.\ Lett.\  B {\bf 683}, 191 (2010)
  [arXiv:0908.3896 [hep-ph]];
  S.~Khalil, M.~U.~Rehman, Q.~Shafi and E.~A.~Zaakouk,
  arXiv:1010.3657 [hep-ph].


\bibitem{Jeannerot:2003qv}
  R.~Jeannerot, J.~Rocher and M.~Sakellariadou,
  Phys.\ Rev.\  D {\bf 68}, 103514 (2003)
  [arXiv:hep-ph/0308134].

\bibitem{Rocher:2004et}
  J.~Rocher and M.~Sakellariadou,
  JCAP {\bf 0503}, 004 (2005)
  [arXiv:hep-ph/0406120].

\bibitem{Bouchet:2000hd}
  F.~R.~Bouchet, P.~Peter, A.~Riazuelo and M.~Sakellariadou,
  Phys.\ Rev.\  D {\bf 65}, 021301 (2002)
  [arXiv:astro-ph/0005022];
  M.~Wyman, L.~Pogosian and I.~Wasserman,
  Phys.\ Rev.\  D {\bf 72}, 023513 (2005)
  [Erratum-ibid.\  D {\bf 73}, 089905 (2006)]
  [arXiv:astro-ph/0503364].

\bibitem{Endo:2003fr}
  M.~Endo, M.~Kawasaki and T.~Moroi,
  Phys.\ Lett.\  B {\bf 569}, 73 (2003)
  [arXiv:hep-ph/0304126].


\bibitem{Peiris:2003ff}
  H.~V.~Peiris {\it et al.}  [WMAP Collaboration],
  Astrophys.\ J.\ Suppl.\  {\bf 148}, 213 (2003)
  [arXiv:astro-ph/0302225].

\bibitem{Dunkley:2010ge}
  J.~Dunkley {\it et al.},
  arXiv:1009.0866 [astro-ph.CO].

\bibitem{Kawasaki:2003zv}
  M.~Kawasaki, M.~Yamaguchi and J.~Yokoyama,
  Phys.\ Rev.\  D {\bf 68}, 023508 (2003)
  [arXiv:hep-ph/0304161];
  M.~Yamaguchi and J.~Yokoyama,
  Phys.\ Rev.\  D {\bf 68}, 123520 (2003)
  [arXiv:hep-ph/0307373];
  M.~Yamaguchi and J.~Yokoyama,
  Phys.\ Rev.\  D {\bf 70}, 023513 (2004)
  [arXiv:hep-ph/0402282].

\bibitem{Feng:2003mk}
  B.~Feng, M.~z.~Li, R.~J.~Zhang and X.~m.~Zhang,
  Phys.\ Rev.\  D {\bf 68}, 103511 (2003)
  [arXiv:astro-ph/0302479];
  D.~J.~H.~Chung, G.~Shiu and M.~Trodden,
  Phys.\ Rev.\  D {\bf 68}, 063501 (2003)
  [arXiv:astro-ph/0305193];
  G.~Ballesteros, J.~A.~Casas and J.~R.~Espinosa,
  JCAP {\bf 0603}, 001 (2006)
  [arXiv:hep-ph/0601134];
  T.~Kobayashi and F.~Takahashi,
  arXiv:1011.3988 [astro-ph.CO].

\bibitem{Covi:2004tp}
  L.~Covi, D.~H.~Lyth, A.~Melchiorri and C.~J.~Odman,
  Phys.\ Rev.\  D {\bf 70}, 123521 (2004)
  [arXiv:astro-ph/0408129].



\bibitem{Lazarides:1997vv}
  G.~Lazarides and N.~D.~Vlachos,
  Phys.\ Rev.\  D {\bf 56}, 4562 (1997)
  [arXiv:hep-ph/9707296];
  N.~Tetradis,
  Phys.\ Rev.\  D {\bf 57}, 5997 (1998)
  [arXiv:astro-ph/9707214];
  L.~E.~Mendes and A.~R.~Liddle,
  Phys.\ Rev.\  D {\bf 62}, 103511 (2000)
  [arXiv:astro-ph/0006020].

\bibitem{Clesse:2009ur}
  S.~Clesse, C.~Ringeval and J.~Rocher,
  Phys.\ Rev.\  D {\bf 80}, 123534 (2009)
  [arXiv:0909.0402 [astro-ph.CO]].



\bibitem{Panagiotakopoulos:1997if}
  C.~Panagiotakopoulos and N.~Tetradis,
  Phys.\ Rev.\  D {\bf 59}, 083502 (1999)
  [arXiv:hep-ph/9710526];
  G.~Lazarides and N.~Tetradis,
  Phys.\ Rev.\  D {\bf 58}, 123502 (1998)
  [arXiv:hep-ph/9802242];
  J.~Lesgourgues,
  Phys.\ Lett.\  B {\bf 452}, 15 (1999)
  [arXiv:hep-ph/9811255];
  J.~Lesgourgues,
  Nucl.\ Phys.\  B {\bf 582}, 593 (2000)
  [arXiv:hep-ph/9911447];
  C.~Panagiotakopoulos,
  Phys.\ Rev.\  D {\bf 71}, 063516 (2005)
  [arXiv:hep-ph/0411143].

\bibitem{Langlois:2004nn}
D.~Langlois and F.~Vernizzi,
Phys.\ Rev.\ D {\bf 70}, 063522 (2004)
[arXiv:astro-ph/0403258];
  T.~Moroi, T.~Takahashi and Y.~Toyoda,
  Phys.\ Rev.\  D {\bf 72}, 023502 (2005)
  [arXiv:hep-ph/0501007];
  T.~Moroi and T.~Takahashi,
  Phys.\ Rev.\  D {\bf 72}, 023505 (2005)
  [arXiv:astro-ph/0505339];
  K.~Ichikawa, T.~Suyama, T.~Takahashi and M.~Yamaguchi,
  Phys.\ Rev.\  D {\bf 78}, 023513 (2008)
  [arXiv:0802.4138 [astro-ph]];
  K.~Ichikawa, T.~Suyama, T.~Takahashi and M.~Yamaguchi,
  Phys.\ Rev.\  D {\bf 78}, 063545 (2008)
  [arXiv:0807.3988 [astro-ph]];
  T.~Suyama, T.~Takahashi, M.~Yamaguchi and S.~Yokoyama,
  arXiv:1009.1979 [astro-ph.CO].

\bibitem{Goncharov:1983mw}
  A.~B.~Goncharov and A.~D.~Linde,
  Phys.\ Lett.\  B {\bf 139}, 27 (1984);
  A.~S.~Goncharov and A.~D.~Linde,
  Class.\ Quant.\ Grav.\  {\bf 1}, L75 (1984);
  H.~Murayama, H.~Suzuki, T.~Yanagida and J.~Yokoyama,
  Phys.\ Rev.\  D {\bf 50}, 2356 (1994)
  [arXiv:hep-ph/9311326].

\bibitem{Kawasaki:2000yn}
  M.~Kawasaki, M.~Yamaguchi and T.~Yanagida,
  Phys.\ Rev.\ Lett.\  {\bf 85}, 3572 (2000)
  [arXiv:hep-ph/0004243];
  M.~Kawasaki, M.~Yamaguchi and T.~Yanagida,
  Phys.\ Rev.\  D {\bf 63}, 103514 (2001)
  [arXiv:hep-ph/0011104].

\bibitem{tHooft} 
G. 't Hooft, in {\it Recent developments in gauge theories}, edited by
G. 't Hooft {\it et al.} (Plenum Press, Carg\`ese, 1980).

\bibitem{Silverstein:2008sg}
  E.~Silverstein and A.~Westphal,
  Phys.\ Rev.\  D {\bf 78}, 106003 (2008)
  [arXiv:0803.3085 [hep-th]];
  L.~McAllister, E.~Silverstein and A.~Westphal,
  arXiv:0808.0706 [hep-th].

\bibitem{Takahashi:2010ky}
  F.~Takahashi,
  Phys.\ Lett.\  B {\bf 693}, 140 (2010)
  [arXiv:1006.2801 [hep-ph]];
  K.~Nakayama and F.~Takahashi,
  JCAP {\bf 1011}, 009 (2010)
  [arXiv:1008.2956 [hep-ph]].

\bibitem{Kallosh:2010xz}
  R.~Kallosh, A.~Linde and T.~Rube,
  arXiv:1011.5945 [hep-th].


\bibitem{Linde:1994hy}
  A.~D.~Linde,
  Phys.\ Lett.\  B {\bf 327}, 208 (1994)
  [arXiv:astro-ph/9402031];
  A.~D.~Linde and D.~A.~Linde,
  Phys.\ Rev.\  D {\bf 50}, 2456 (1994)
  [arXiv:hep-th/9402115];
  A.~Vilenkin,
  Phys.\ Rev.\ Lett.\  {\bf 72}, 3137 (1994)
  [arXiv:hep-th/9402085].


\bibitem{Sakai:1995nh}
  N.~Sakai, H.~A.~Shinkai, T.~Tachizawa and K.~i.~Maeda,
  Phys.\ Rev.\  D {\bf 53}, 655 (1996)
  [Erratum-ibid.\  D {\bf 54}, 2981 (1996)]
  [Phys.\ Rev.\  D {\bf 54}, 2981 (1996)]
  [arXiv:gr-qc/9506068];
  N.~Sakai,
  Phys.\ Rev.\  D {\bf 54}, 1548 (1996)
  [arXiv:gr-qc/9512045].

\bibitem{Kawasaki:2000tv}
  M.~Kawasaki, N.~Sakai, M.~Yamaguchi and T.~Yanagida,
  Phys.\ Rev.\  D {\bf 62}, 123507 (2000)
  [arXiv:hep-ph/0005073].

\bibitem{Izawa:1998rh}
  K.~I.~Izawa, M.~Kawasaki and T.~Yanagida,
  Prog.\ Theor.\ Phys.\  {\bf 101}, 1129 (1999)
  [arXiv:hep-ph/9810537].

\bibitem{Kawasaki:2001as}
  M.~Kawasaki and M.~Yamaguchi,
  Phys.\ Rev.\  D {\bf 65}, 103518 (2002)
  [arXiv:hep-ph/0112093].


\bibitem{Freese:1990rb}
  K.~Freese, J.~A.~Frieman and A.~V.~Olinto,
  Phys.\ Rev.\ Lett.\  {\bf 65}, 3233 (1990);
  F.~C.~Adams, J.~R.~Bond, K.~Freese, J.~A.~Frieman and A.~V.~Olinto,
  Phys.\ Rev.\  D {\bf 47}, 426 (1993)
  [arXiv:hep-ph/9207245].

\bibitem{German:2001sm}
  G.~German, A.~Mazumdar and A.~Perez-Lorenzana,
  Mod.\ Phys.\ Lett.\  A {\bf 17}, 1627 (2002)
  [arXiv:hep-ph/0111371].


\bibitem{Binetruy:1996xj}
  P.~Binetruy and G.~R.~Dvali,
  Phys.\ Lett.\  B {\bf 388}, 241 (1996)
  [arXiv:hep-ph/9606342].

\bibitem{Halyo:1996pp}
  E.~Halyo,
  Phys.\ Lett.\  B {\bf 387}, 43 (1996)
  [arXiv:hep-ph/9606423].

\bibitem{Binetruy:2004hh}
  P.~Binetruy, G.~Dvali, R.~Kallosh and A.~Van Proeyen,
  Class.\ Quant.\ Grav.\  {\bf 21}, 3137 (2004)
  [arXiv:hep-th/0402046].

\bibitem{Rocher:2004my}
  J.~Rocher and M.~Sakellariadou,
  Phys.\ Rev.\ Lett.\  {\bf 94}, 011303 (2005)
  [arXiv:hep-ph/0412143].


\bibitem{Battye:2010xz}
  R.~Battye and A.~Moss,
  Phys.\ Rev.\  D {\bf 82}, 023521 (2010)
  [arXiv:1005.0479 [astro-ph.CO]].


\bibitem{Seto:2005qg}
  O.~Seto and J.~Yokoyama,
  Phys.\ Rev.\  D {\bf 73}, 023508 (2006)
  [arXiv:hep-ph/0508172];
  J.~Rocher and M.~Sakellariadou,
  JCAP {\bf 0611}, 001 (2006)
  [arXiv:hep-th/0607226].



\bibitem{Kadota:2007nc}
  K.~Kadota and M.~Yamaguchi,
  Phys.\ Rev.\  D {\bf 76}, 103522 (2007)
  [arXiv:0706.2676 [hep-ph]].

\bibitem{Kadota:2008pm}
  K.~Kadota, T.~Kawano and M.~Yamaguchi,
  Phys.\ Rev.\  D {\bf 77}, 123516 (2008)
  [arXiv:0802.0525 [hep-ph]].

\bibitem{Yamaguchi:2005qm}
  M.~Yamaguchi and J.~Yokoyama,
  Phys.\ Rev.\  D {\bf 74}, 043523 (2006)
  [arXiv:hep-ph/0512318].

\bibitem{Kawano:2007gg}
  T.~Kawano,
  Prog.\ Theor.\ Phys.\  {\bf 120}, 793 (2008)
  [arXiv:0712.2351 [hep-th]];
  T.~Kawano and M.~Yamaguchi,
  Phys.\ Rev.\  D {\bf 78}, 123511 (2008)
  [arXiv:0806.4971 [hep-th]].


\bibitem{Fukugita:1986hr}
  M.~Fukugita and T.~Yanagida,
  Phys.\ Lett.\  B {\bf 174}, 45 (1986).

\bibitem{Einhorn:2009bh}
  M.~B.~Einhorn and D.~R.~T.~Jones,
  JHEP {\bf 1003}, 026 (2010)
  [arXiv:0912.2718 [hep-ph]].

\bibitem{Lee:2010hj}
  H.~M.~Lee,
  JCAP {\bf 1008}, 003 (2010)
  [arXiv:1005.2735 [hep-ph]].

\bibitem{Ferrara:2010yw}
  S.~Ferrara, R.~Kallosh, A.~Linde, A.~Marrani and A.~Van Proeyen,
  Phys.\ Rev.\  D {\bf 82}, 045003 (2010)
  [arXiv:1004.0712 [hep-th]];
  S.~Ferrara, R.~Kallosh, A.~Linde, A.~Marrani and A.~Van Proeyen,
  arXiv:1008.2942 [hep-th].

\bibitem{Nakayama:2010ga}
  K.~Nakayama and F.~Takahashi,
  JCAP {\bf 1011}, 039 (2010)
  [arXiv:1009.3399 [hep-ph]].


\bibitem{Chiba:2008ia}
  T.~Chiba and M.~Yamaguchi,
  JCAP {\bf 0810}, 021 (2008)
  [arXiv:0807.4965 [astro-ph]];
  T.~Chiba and M.~Yamaguchi,
  JCAP {\bf 0901}, 019 (2009)
  [arXiv:0810.5387 [astro-ph]].


\bibitem{Makino:1991sg}
  N.~Makino and M.~Sasaki,
  Prog.\ Theor.\ Phys.\  {\bf 86}, 103 (1991);
  R.~Fakir, S.~Habib and W.~Unruh,
  Astrophys.\ J.\  {\bf 394}, 396 (1992).
  J.~c.~Hwang,
  Class.\ Quant.\ Grav.\  {\bf 14}, 1981 (1997)
  [arXiv:gr-qc/9605024].
  E.~Komatsu and T.~Futamase,
  Phys.\ Rev.\  D {\bf 59}, 064029 (1999)
  [arXiv:astro-ph/9901127].


\bibitem{Futamase:1987ua}
  T.~Futamase and K.~i.~Maeda,
  Phys.\ Rev.\  D {\bf 39}, 399 (1989);


\bibitem{DeSimone:2008ei}
  A.~De Simone, M.~P.~Hertzberg and F.~Wilczek,
  Phys.\ Lett.\  B {\bf 678}, 1 (2009)
  [arXiv:0812.4946 [hep-ph]];
  F.~L.~Bezrukov, A.~Magnin and M.~Shaposhnikov,
  Phys.\ Lett.\  B {\bf 675}, 88 (2009)
  [arXiv:0812.4950 [hep-ph]];
  J.~L.~F.~Barbon and J.~R.~Espinosa,
  Phys.\ Rev.\  D {\bf 79}, 081302 (2009)
  [arXiv:0903.0355 [hep-ph]];
  F.~Bezrukov and M.~Shaposhnikov,
  JHEP {\bf 0907}, 089 (2009)
  [arXiv:0904.1537 [hep-ph]];
  A.~O.~Barvinsky, A.~Y.~Kamenshchik, C.~Kiefer, A.~A.~Starobinsky and C.~Steinwachs,
  JCAP {\bf 0912}, 003 (2009)
  [arXiv:0904.1698 [hep-ph]];
  A.~O.~Barvinsky, A.~Y.~Kamenshchik, C.~Kiefer, A.~A.~Starobinsky and C.~F.~Steinwachs,
  arXiv:0910.1041 [hep-ph];
  C.~P.~Burgess, H.~M.~Lee and M.~Trott,
  JHEP {\bf 1007}, 007 (2010)
  [arXiv:1002.2730 [hep-ph]];
  R.~N.~Lerner and J.~McDonald,
  Phys.\ Rev.\  D {\bf 82}, 103525 (2010)
  [arXiv:1005.2978 [hep-ph]];
  F.~Bezrukov, A.~Magnin, M.~Shaposhnikov and S.~Sibiryakov,
  arXiv:1008.5157 [hep-ph].



\bibitem{Kallosh:2010ug}
  R.~Kallosh and A.~Linde,
  JCAP {\bf 1011}, 011 (2010)
  [arXiv:1008.3375 [hep-th]].


\bibitem{Kaneda:2010ut}
  S.~Kaneda, S.~V.~Ketov and N.~Watanabe,
  Mod.\ Phys.\ Lett.\  A {\bf 25}, 2753 (2010)
  [arXiv:1001.5118 [hep-th]];
  S.~Kaneda, S.~V.~Ketov and N.~Watanabe,
  Class.\ Quant.\ Grav.\  {\bf 27}, 145016 (2010)
  [arXiv:1002.3659 [hep-th]];
  S.~V.~Ketov,
  Phys.\ Lett.\  B {\bf 692}, 272 (2010)
  [arXiv:1005.3630 [hep-th]];
  S.~V.~Ketov and A.~A.~Starobinsky,
  arXiv:1011.0240 [hep-th].

\bibitem{Gates:2009hu}
  S.~J.~J.~Gates and S.~V.~Ketov,
  Phys.\ Lett.\  B {\bf 674}, 59 (2009)
  [arXiv:0901.2467 [hep-th]];
  S.~V.~Ketov,
  Class.\ Quant.\ Grav.\  {\bf 26}, 135006 (2009)
  [arXiv:0903.0251 [hep-th]].


\bibitem{ArmendarizPicon:1999rj}
  C.~Armendariz-Picon, T.~Damour and V.~F.~Mukhanov,
  Phys.\ Lett.\  B {\bf 458}, 209 (1999)
  [arXiv:hep-th/9904075].

\bibitem{ArkaniHamed:2003uz}
  N.~Arkani-Hamed, P.~Creminelli, S.~Mukohyama and M.~Zaldarriaga,
  JCAP {\bf 0404}, 001 (2004)
  [arXiv:hep-th/0312100].

\bibitem{Alishahiha:2004eh}
  M.~Alishahiha, E.~Silverstein and D.~Tong,
  Phys.\ Rev.\  D {\bf 70}, 123505 (2004)
  [arXiv:hep-th/0404084];
  E.~Silverstein and D.~Tong,
  Phys.\ Rev.\  D {\bf 70}, 103505 (2004)
  [arXiv:hep-th/0310221].


\bibitem{Kobayashi:2010cm}
  T.~Kobayashi, M.~Yamaguchi and J.~Yokoyama,
  Phys.\ Rev.\ Lett.\  {\bf 105}, 231302 (2010)
  [arXiv:1008.0603 [hep-th]].

\bibitem{Khoury:2010gb}
  J.~Khoury, J.~L.~Lehners and B.~Ovrut,
  arXiv:1012.3748 [hep-th].














\end{thebibliography}
\end{document}